\pdfoutput=1
\documentclass[letterpaper,twocolumn,10pt]{article}
\pdftrailerid{}
\usepackage[letterpaper,top=1in,bottom=1in,left=0.75in,right=0.75in,columnsep=0.33in]{geometry}
\usepackage{mathptmx}
\usepackage[T1]{fontenc}
\usepackage[utf8]{inputenc}
\usepackage{pslatex}
\usepackage[kerning,spacing]{microtype}
\usepackage{cite}
\usepackage{url}
\usepackage{hyperref}
\usepackage{relsize}
\usepackage{fontawesome5}
\usepackage[most]{tcolorbox}
\usepackage{amsmath,amssymb,amsfonts}
\usepackage{graphicx}
\usepackage{textcomp}
\usepackage{enumitem}
\usepackage{tikz}
\usetikzlibrary{arrows.meta, shadows.blur, positioning, calc, fit, backgrounds,
  shapes.geometric, shapes.misc, decorations.pathreplacing, patterns, matrix}
\usepackage{pgfplots}
\pgfplotsset{compat=1.18}
\usepgfplotslibrary{fillbetween, groupplots}
\usepackage{xcolor}
\definecolor{okblue}{HTML}{0072B2}
\definecolor{okorange}{HTML}{E69F00}
\definecolor{okverm}{HTML}{D55E00}
\definecolor{okgreen}{HTML}{009E73}
\definecolor{oksky}{HTML}{56B4E9}
\definecolor{okyellow}{HTML}{F0E442}
\definecolor{okpurple}{HTML}{CC79A7}
\definecolor{okgray}{HTML}{6E6E6E}
\usepackage{colortbl}
\usepackage{booktabs}
\usepackage{multirow}
\usepackage{makecell}
\usepackage{pifont}
\usepackage{array}
\usepackage{tabularx}
\usepackage{wasysym}
\usepackage{adjustbox}
\usepackage{caption}
\usepackage{ragged2e}
\usepackage{xurl}
\hypersetup{colorlinks=true, linkcolor=blue, citecolor=blue, urlcolor=blue, pdfauthor={}, pdfcreator={}, pdfproducer={}}

\newcommand{\rih}[1]{\par\vspace{1pt}\noindent\textbf{#1}\space}
\newcommand{\cmark}{\ding{51}}
\newcommand{\xmark}{\ding{55}}
\DeclareRobustCommand{\fullc}{\tikz[baseline=-0.32ex]{\fill (0,0) circle ({0.33em+0.2pt});}}
\DeclareRobustCommand{\halfc}{\tikz[baseline=-0.32ex]{\draw[line width=0.4pt] (0,0) circle (0.33em);\fill (0,0.33em) arc (90:270:0.33em) -- cycle;}}
\DeclareRobustCommand{\capbox}[2]{{\setlength{\fboxsep}{1pt}\colorbox{#1}{\vphantom{Ag}#2}}}
\DeclareRobustCommand{\emptyc}{\tikz[baseline=-0.32ex]{\draw[line width=0.4pt] (0,0) circle (0.33em);}}

\newcommand{\nSchemes}{26}%
\newcommand{\nAcademic}{2}%
\newcommand{\nBroker}{8}%
\newcommand{\nOSS}{16}%
\newcommand{\nExamined}{28}%
\newcommand{\nExcludedReview}{2}%
\newcommand{\nAddedSchemes}{13}%
\newcommand{\nAtwo}{3}%
\newcommand{\nAthree}{7}%
\newcommand{\nAfour}{10}%
\newcommand{\nAutoExec}{17}%
\newcommand{\nWfour}{14}%
\newcommand{\nHassignment}{6}%
\newcommand{\nHunbounded}{13}%
\newcommand{\nUnboundedAuto}{10}%
\newcommand{\nHtwoPlatform}{15}%
\newcommand{\nPlatformUnbounded}{11}%
\newcommand{\nHthreeInterceptable}{9}%
\newcommand{\nRoneFull}{4}%
\newcommand{\nRtwoFull}{3}%
\newcommand{\nRthreeFull}{3}%
\newcommand{\nRfourFull}{3}%
\newcommand{\nRfiveFull}{0}%
\newcommand{\nRsixFull}{0}%
\newcommand{\nRsevenFull}{3}%
\newcommand{\nConeFull}{0}%
\newcommand{\nConePartial}{8}%
\newcommand{\nCtwoFull}{0}%
\newcommand{\nCthreeFull}{2}%
\newcommand{\nCfourFull}{0}%
\newcommand{\nCfourPartial}{5}%
\newcommand{\nCfiveFull}{0}%
\newcommand{\nCsixFull}{1}%
\newcommand{\nCsevenFull}{1}%
\newcommand{\nCeightFull}{10}%
\newcommand{\nCeightPartial}{7}%
\newcommand{\nCnineFull}{3}%
\newcommand{\nCtenFull}{5}%
\newcommand{\nCelevenFull}{0}%
\newcommand{\nCelevenPartial}{4}%
\newcommand{\nInputVerified}{0}%
\newcommand{\nOrderCapable}{22}%
\newcommand{\nBrokerAutoExec}{6}%
\newcommand{\nBrokerUnbounded}{6}%
\newcommand{\nBrokerCeightFull}{3}%
\newcommand{\nBrokerCtwoNone}{8}%
\newcommand{\nOSSAutoExec}{11}%
\newcommand{\nOSSUnbounded}{7}%
\newcommand{\nOSSAdopted}{7}%
\newcommand{\nOSSPurposive}{9}%
\newcommand{\nCensus}{17}%
\newcommand{\nCensusUnbounded}{11}%
\newcommand{\starsOpenalgo}{2{,}722}%
\newcommand{\starsLumibot}{2{,}086}%
\newcommand{\nUnion}{765}%
\newcommand{\nQueryRaw}{1{,}232}%
\newcommand{\nStageOneKept}{468}%
\newcommand{\nEligible}{215}%
\newcommand{\nEligibleZeroStar}{161}%
\newcommand{\nEligibleTwentySix}{198}%
\newcommand{\nEligibleTwoWeek}{129}%
\newcommand{\nEligibleTwentyStars}{6}%
\newcommand{\nExcludeEone}{11}%
\newcommand{\nExcludeEtwo}{52}%
\newcommand{\nExcludeEthree}{24}%
\newcommand{\nExcludeEfour}{0}%
\newcommand{\nExcludeEfive}{114}%
\newcommand{\nExcludeEsix}{46}%
\newcommand{\nExcludeEseven}{6}%
\newcommand{\nEligibleUnscored}{201}%
\newcommand{\starThreshold}{20}%
\newcommand{\nextStarBelow}{11}%
\newcommand{\nEligibleBeforeTwentySix}{17}%
\newcommand{\nEligibleHackathon}{87}%
\newcommand{\nEligibleAlpaca}{136}%
\newcommand{\nEligibleTwoWeekAlpaca}{119}%
\newcommand{\pctEligibleTwoWeek}{60.0\%}%
\newcommand{\pctEligibleZeroStar}{74.9\%}%
\newcommand{\timelineBefore}{10}%
\newcommand{\nUnboundedNoLimit}{5}%
\newcommand{\nBrokerUnboundedNoLimit}{3}%
\newcommand{\nOrderNoFullC}{20}%
\newcommand{\nCeightFullUnbounded}{4}%
\newcommand{\nRoneFullOrder}{4}%
\newcommand{\nOSSAdoptedCeightFull}{1}%
\newcommand{\nPurposiveOrderCapable}{7}%
\newcommand{\nPurposiveCeightFull}{6}%
\newcommand{\nMentionMCP}{135}%
\newcommand{\nMentionClaude}{102}%
\newcommand{\nMentionOpenAI}{63}%
\newcommand{\nMentionGemini}{35}%
\newcommand{\nMentionOpenWeight}{46}%
\newcommand{\nMentionGreeksIV}{120}%
\newcommand{\nMentionSpreads}{64}%
\newcommand{\nMentionWriting}{22}%
\newcommand{\nMentionMultiVendor}{58}%

\definecolor{paired-light-blue}{RGB}{198, 219, 239}
\definecolor{paired-dark-blue}{RGB}{49, 130, 188}
\definecolor{paired-light-orange}{RGB}{251, 208, 162}
\definecolor{paired-dark-orange}{RGB}{230, 85, 12}
\definecolor{paired-light-green}{RGB}{199, 233, 193}
\definecolor{paired-dark-green}{RGB}{49, 163, 83}
\definecolor{paired-light-purple}{RGB}{218, 218, 235}
\definecolor{paired-dark-purple}{RGB}{117, 107, 176}
\definecolor{paired-light-gray}{RGB}{217, 217, 217}
\definecolor{paired-dark-gray}{RGB}{99, 99, 99}
\definecolor{paired-light-red}{RGB}{252, 187, 161}
\definecolor{paired-dark-red}{RGB}{227, 74, 51}

\newtcolorbox{takeaway}{enhanced, colback=okblue!5, colframe=okblue!55, boxrule=0.5pt, arc=1.5pt,
  left=3pt, right=3pt, top=2pt, bottom=2pt, before skip=4pt, after skip=4pt}

\begin{document}

\date{}
\title{\Large \bf Unbounded Losses, Unchecked Inputs: Measuring the\\ Safeguards of LLM Options-Trading Agents}

\author{
{\rm Mengxiao Wang}\\
Texas A\&M University\\
\texttt{jacksonwang@tamu.edu}
\and
{\rm Nitesh Saxena}\\
Texas A\&M University\\
\texttt{nsaxena@tamu.edu}
}

\maketitle

\begin{abstract}
Large language model (LLM) agents now place options orders at retail brokers, and developers publish open-source options agents faster than anyone has examined them. Prior security work on trading agents scored academic equities designs by whether an attacker can flip a buy-or-sell decision. We present the first measurement of the options agents that brokers and developers ship, scored by the safeguards they enforce and by the worst loss their documented order paths can reach. We screened \nUnion{} GitHub repositories, broker documentation, and the research literature and found \nEligible{} eligible repositories and \nBroker{} broker-native agent interfaces. We then scored \nSchemes{} systems, including every broker interface and every open-source project with at least \starThreshold{} stars, on seven robustness and eleven security safeguards with cell-level evidence. The ecosystem is new: \nEligibleTwentySix{} of the \nEligible{} repositories were created in 2026, and \nEligibleTwoWeek{} (\pctEligibleTwoWeek{}) appeared in the two weeks around one options-trading hackathon. Of the \nOrderCapable{} systems that can place orders, \nCeightFull{} enforce a pre-trade limit outside the model, yet none of the \nSchemes{} authenticate their text sources or cross-check market data. Of the \nSchemes{}, \nAutoExec{} can execute with no mandatory human step and \nHunbounded{} can reach unbounded loss, including \nBrokerUnbounded{} of the \nBroker{} broker interfaces. Because an attacker who writes the agent's inputs can move an order's strike, expiry, volatility, and legs while its direction stays correct, options agents need loss limits enforced at the order boundary.
\end{abstract}

\section{Introduction}
\label{sec:intro}

\begin{figure*}[t]
\centering
\includegraphics[width=\textwidth]{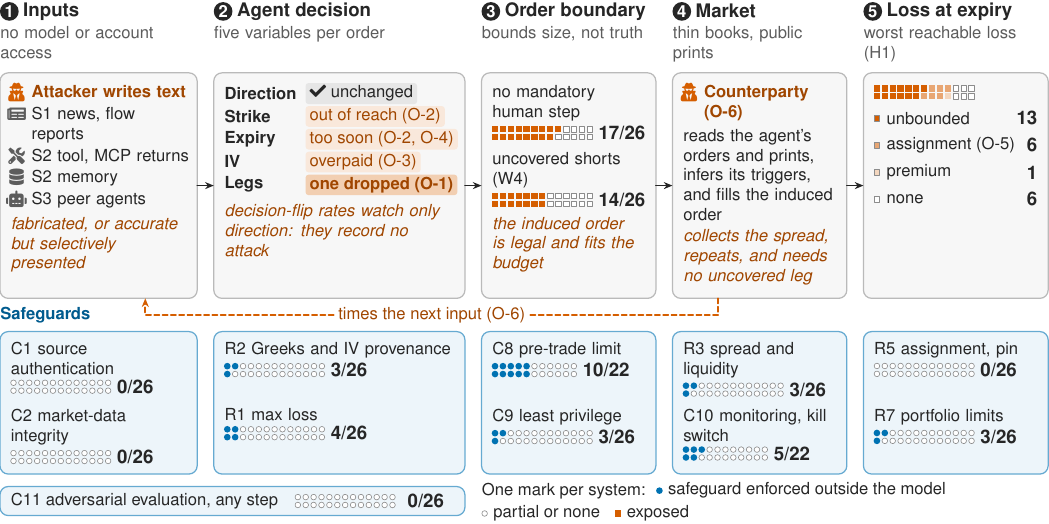}
\caption{\textbf{Attack path and the safeguards we measured.} Top: an attacker who writes text into the agent's inputs (S1--S3) moves four of the five decisions of an options order while its direction stays correct (O-1 to O-4). The induced order passes limits that bound its size, a counterparty who reads public prints can fill it (O-6), and the position can end in assignment (O-5) or unbounded loss. Bottom: two safeguards per step, one mark per scored system, filled if enforced outside the model. C8, the most common safeguard, is enforced by \nCeightFull{} of the \nOrderCapable{} order-capable systems (Table~\ref{tab:schemes}).}
\label{fig:overview}
\end{figure*}

A large language model (LLM) agent can now open an options position at a retail broker with no human touching the order. Robinhood announced that agentic options trading is available to all US customers~\cite{robinhoodoptionsallcustomers2026,robinhoodoptions2026}, on a platform with about 27 million funded customers~\cite{robinhoodq12026}. The tastytrade Model Context Protocol (MCP) server exposes equity, index, and futures options, including multi-leg orders, to agents with write access~\cite{tastytrademcp2026,tastytradenews2026}, and 7 of the 14 brokers reviewed by StockBrokers.com in September 2026 run a first-party agent server~\cite{stockbrokers2026}. Developers are building on these interfaces even faster. Of the \nEligible{} public options-agent repositories we found on GitHub, \nEligibleTwentySix{} were created in 2026 (\S\ref{sec:rq1}). The market they trade is large and short-dated, since listed options averaged 72.8 million contracts a day in 2026 Q2, up more than 19\% from a year earlier~\cite{occvolume2026}, and zero-days-to-expiry (0DTE) contracts reached a record 66.2\% of S\&P 500 index-options volume in July 2026~\cite{cboe2026}.

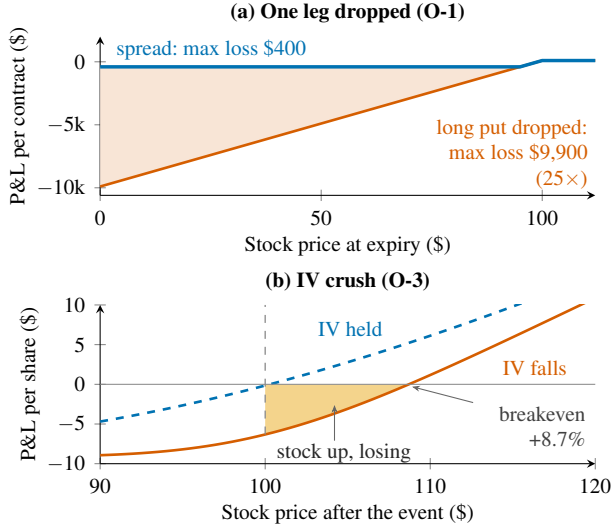
\begin{figure}[!htb]
\centering
\begin{tikzpicture}
\begin{groupplot}[
  group style={group size=1 by 2, vertical sep=1.45cm},
  scale only axis, width=6.55cm, height=2.1cm,
  tick label style={font=\footnotesize}, label style={font=\footnotesize},
  title style={font=\footnotesize\bfseries, yshift=-4pt},
  axis lines=left,
  every axis plot/.append style={line width=1.0pt},
  xlabel style={font=\footnotesize, yshift=3pt}, ylabel style={font=\footnotesize, yshift=-4pt},
]
\nextgroupplot[title={(a) One leg dropped (O-1)}, xlabel={Stock price at expiry (\$)},
  ylabel={P\&L per contract (\$)}, xmin=0, xmax=112, ymin=-10.6, ymax=2.0,
  xtick={0,50,100},
  ytick={-10,-5,0}, yticklabels={$-$10k,$-$5k,0}]
\addplot[name path=NK, color=okverm] coordinates {(0,-9.9) (100,0.1) (112,0.1)};
\addplot[name path=SP, color=okblue, line width=1.3pt] coordinates {(0,-0.4) (95,-0.4) (100,0.1) (112,0.1)};
\addplot[fill=okverm!16, draw=none, forget plot] fill between[of=SP and NK, soft clip={domain=0:95}];
\node[font=\footnotesize, text=okblue, anchor=south west, align=left, inner sep=1pt] at (axis cs:3,-0.1) {spread: max loss \$400};
\node[font=\footnotesize, text=okverm, anchor=south east, align=right, inner sep=1pt] at (axis cs:111,-10.3) {long put dropped:\\ max loss \$9{,}900\\ (25$\times$)};
\nextgroupplot[title={(b) IV crush (O-3)}, xlabel={Stock price after the event (\$)},
  ylabel={P\&L per share (\$)}, xmin=90, xmax=120, ymin=-10, ymax=10,
  xtick={90,100,110,120}, ytick={-10,-5,0,5,10}]
\addplot[name path=CR, color=okverm] coordinates {(90.00,-8.932) (90.50,-8.893) (91.00,-8.848) (91.50,-8.796) (92.00,-8.738) (92.50,-8.671) (93.00,-8.595) (93.50,-8.511) (94.00,-8.416) (94.50,-8.311) (95.00,-8.194) (95.50,-8.066) (96.00,-7.926) (96.50,-7.773) (97.00,-7.607) (97.50,-7.427) (98.00,-7.234) (98.50,-7.026) (99.00,-6.805) (99.50,-6.569) (100.00,-6.319) (100.50,-6.055) (101.00,-5.777) (101.50,-5.485) (102.00,-5.180) (102.50,-4.861) (103.00,-4.530) (103.50,-4.186) (104.00,-3.830) (104.50,-3.463) (105.00,-3.085) (105.50,-2.697) (106.00,-2.299) (106.50,-1.891) (107.00,-1.475) (107.50,-1.051) (108.00,-0.620) (108.50,-0.181) (109.00,0.264) (109.50,0.714) (110.00,1.171) (110.50,1.632) (111.00,2.097) (111.50,2.567) (112.00,3.040) (112.50,3.517) (113.00,3.996) (113.50,4.478) (114.00,4.962) (114.50,5.449) (115.00,5.937) (115.50,6.427) (116.00,6.918) (116.50,7.411) (117.00,7.905) (117.50,8.399) (118.00,8.894) (118.50,9.390) (119.00,9.887) (119.50,10.384) (120.00,10.882)};
\addplot[color=okblue, dashed] coordinates {(90.00,-4.692) (90.50,-4.509) (91.00,-4.322) (91.50,-4.129) (92.00,-3.932) (92.50,-3.731) (93.00,-3.524) (93.50,-3.313) (94.00,-3.098) (94.50,-2.878) (95.00,-2.653) (95.50,-2.423) (96.00,-2.189) (96.50,-1.951) (97.00,-1.707) (97.50,-1.459) (98.00,-1.207) (98.50,-0.950) (99.00,-0.689) (99.50,-0.423) (100.00,-0.153) (100.50,0.122) (101.00,0.401) (101.50,0.684) (102.00,0.972) (102.50,1.264) (103.00,1.560) (103.50,1.860) (104.00,2.164) (104.50,2.473) (105.00,2.785) (105.50,3.102) (106.00,3.422) (106.50,3.746) (107.00,4.075) (107.50,4.407) (108.00,4.742) (108.50,5.082) (109.00,5.425) (109.50,5.771) (110.00,6.121) (110.50,6.475) (111.00,6.832) (111.50,7.193) (112.00,7.556) (112.50,7.923) (113.00,8.294) (113.50,8.667) (114.00,9.044) (114.50,9.423) (115.00,9.806) (115.50,10.191) (116.00,10.580) (116.50,10.971) (117.00,11.365) (117.50,11.762) (118.00,12.161) (118.50,12.563) (119.00,12.968) (119.50,13.375) (120.00,13.784)};
\addplot[name path=ZERO, draw=none, forget plot] coordinates {(90,0) (120,0)};
\addplot[fill=okorange!45, draw=none, forget plot] fill between[of=CR and ZERO, soft clip={domain=100:108.70}];
\draw[black!45, line width=0.4pt] (axis cs:90,0) -- (axis cs:120,0);
\draw[dashed, black!45, line width=0.5pt] (axis cs:100,-6.32) -- (axis cs:100,10);
\node[font=\footnotesize, text=okblue, anchor=south east, inner sep=1pt] at (axis cs:107.3,5.6) {IV held};
\node[font=\footnotesize, text=okverm, anchor=north west, inner sep=1pt] at (axis cs:114.2,3.4) {IV falls};
\node[font=\footnotesize, text=black!85, anchor=north west, inner sep=0.5pt] at (axis cs:100.6,-7.2) {stock up, losing};
\draw[black!60, line width=0.4pt, -{Stealth[length=3pt]}] (axis cs:104.2,-7.1) -- (axis cs:104.2,-2.6);
\draw[black!60, line width=0.4pt, -{Stealth[length=3pt]}] (axis cs:112.6,-2.3) -- (axis cs:108.9,-0.3);
\node[font=\footnotesize, text=black!75, anchor=north east, align=right, inner sep=1pt] at (axis cs:119.6,-2.4) {breakeven\\ +8.7\%};
\end{groupplot}
\end{tikzpicture}
\caption{\textbf{Two ways to lose money with the direction unchanged.} (a)~Structure: a bull put spread (sell the 100 put, buy the 95 put, \$1.00 credit) risks at most \$400 per contract, and the same position without its long put risks \$9{,}900 (O-1). The shaded area is the loss that the dropped leg would have capped. (b)~Volatility: a 30-day at-the-money call bought at 80\% implied volatility ahead of a scheduled event loses money after volatility falls to 25\% unless the stock rises more than 8.7\% (O-3). The shaded band is a correct view that still loses. Black--Scholes values~\cite{blackscholes1973,hull2021} with illustrative parameters.}
\label{fig:payoff}
\end{figure}

An attacker who can publish text that an options agent reads can steer its orders without touching the model or the account, because indirect prompt injection~\cite{greshake2023} and MCP tool poisoning~\cite{mcptox2026} let news, social posts, filings, and tool outputs control an agent's actions. However, the worst case on options keeps the trade's direction. Whereas a stock position reaches unbounded loss through an explicit short~\cite{secshortsale}, the one variable that equities risk checks already watch, an options position reaches it when one leg is dropped while the direction and the named strategy stay the same. If text frames the protective leg of a bull put spread as a drag on returns and the agent drops it, a position that risked \$400 per contract becomes an uncovered put that can lose \$9{,}900, about 25 times more (Figure~\ref{fig:payoff}). Human-run funds have failed this way. In 2018, a fund that sold uncovered natural-gas options told about 290 clients that their balances were gone and that they could owe their clearing broker more, after an 18\% price spike~\cite{optionsellers2018cnbc}.

Despite this exposure, no one has measured whether the agents in use today can stop such an order. The closest work, the FARSIGHT systematization of Wang and Saxena~\cite{wang2026soktradingagentsmarket}, scored 15 academic equities designs by whether an attacker can flip a buy-or-sell decision. Agent-security benchmarks~\cite{debenedetti2024agentdojo,zhan2024injecagent,zhang2025asb} count unintended tool calls, and attack studies of trading agents~\cite{tradetrap2025,marketsignalinjection2026} target equities and pricing agents. Because none of them examines the options agents that brokers and developers ship, or prices a failure by the loss the resulting order can reach, we ask three research questions.
\begin{itemize}[leftmargin=*,itemsep=1pt,topsep=2pt]
  \item \textbf{RQ1 (Ecosystem).} Who builds LLM options-trading agents, and how fast is that population growing?
  \item \textbf{RQ2 (Exposure).} What can these agents do to an account without a human, and what is the worst loss their documented order paths can reach?
  \item \textbf{RQ3 (Safeguards).} Which robustness and security safeguards do they enforce outside the model, and do any check the inputs an attacker writes?
\end{itemize}

To answer them, we screened all \nUnion{} repositories returned by six GitHub searches, read broker documentation and the research literature, and scored \nSchemes{} systems at a single snapshot (\S\ref{sec:method}). The scored set contains every broker-native agent interface (\nBroker{}), every academic system (\nAcademic{}), and every open-source project with at least \starThreshold{} stars (\nOSSAdopted{}) that we found, plus \nOSSPurposive{} lower-star repositories chosen for design coverage. Our codebook grades seven robustness and eleven security safeguards as enforced outside the model, dependent on the model, or undocumented, and every cell cites the public artifact behind it.

The measurement shows a young population whose controls bound the size of an order and leave its inputs unchecked. First (RQ1), \nEligibleTwoWeek{} (\pctEligibleTwoWeek{}) of the \nEligible{} eligible repositories appeared in the two weeks around one options-trading hackathon, \nEligibleAlpaca{} name a single broker, and \nEligibleZeroStar{} have no stars. Second (RQ2), \nAutoExec{} of the \nSchemes{} systems execute with no mandatory human step, \nWfour{} can open an uncovered short, and \nHunbounded{} can reach unbounded loss, among them \nBrokerUnbounded{} of the \nBroker{} broker interfaces. Third (RQ3), \nCeightFull{} of the \nOrderCapable{} order-capable systems enforce a pre-trade limit outside the model, none of the \nSchemes{} authenticate their sources or cross-check market data, and \nUnboundedNoLimit{} can reach unbounded loss with neither control.

We make the following contributions:
\begin{itemize}[leftmargin=*,itemsep=1pt,topsep=2pt]
  \item We give the first measurement of LLM options-trading agents, from a logged screen of \nUnion{} GitHub repositories to \nSchemes{} scored systems, and show that most of the open-source population appeared around a single hackathon (\S\ref{sec:method}, \S\ref{sec:rq1}).
  \item We define a threat model in which the attacker keeps the trade's direction and moves its strike, expiry, volatility, or legs, and derive six option-only attack primitives, including a counterparty attack that survives a ban on uncovered writing (\S\ref{sec:threat}).
  \item We score every system on eighteen safeguards with cell-level evidence and show that the worst case is reachable on broker interfaces, while no system verifies its inputs (\S\ref{sec:rq2}, \S\ref{sec:rq3}).
  \item We derive order-boundary controls for brokers and agent builders that work without detecting manipulated text, and we state five open problems for researchers (\S\ref{sec:recommend}, \S\ref{sec:open}).
\end{itemize}

Because of this measurement, a broker that opens options to agents can check its order path against three controls that work without detecting manipulated text: defined-risk structures for agent orders, a maximum loss computed and enforced before entry, which \nRoneFullOrder{} of the \nOrderCapable{} order-capable systems have today, and a pre-trade limit enforced outside the model, which \nCeightFull{} have.

\section{Background}
\label{sec:background}

Options matter for agent security because an option's loss is no fixed multiple of its price move. A US equity option gives its holder the right to buy (a call) or sell (a put) 100 shares of an underlying at a fixed strike on or before an expiry date, and the holder may also let it lapse~\cite{hull2021}. A long stock position pays off linearly and loses at most the capital committed. Options break both properties.

\rih{Payoff asymmetry and defined-risk structures.}
The option buyer's maximum loss is the premium paid. The seller collects that premium and takes on the obligation, so a short put loses at most the strike minus the premium, while a short call written without an offsetting position, a \emph{naked} call, has no upper bound because the underlying price has none~\cite{hull2021}. Traders cap that exposure by pairing legs. A vertical spread sells one option and buys a further-out-of-the-money option of the same expiry, so its maximum loss is the strike width minus the net credit, and a covered call or cash-secured put writes against shares or cash already held. Thus, the bound comes from the joint configuration of the position. When the protective leg is dropped, or the option is written uncovered, a defined-risk position becomes a naked one with the directional view unchanged. Equities have no such transition that keeps the direction, and it underlies the worst case of \S\ref{sec:primitives}.

\rih{Short selling, margin, and expiry.}
Unbounded loss exists on equities too, since a short sale has no loss ceiling~\cite{secshortsale} and Regulation T limits initial leverage on stock to 2:1~\cite{regt220_12}. The difference is where the decision sits. While a stock position becomes unbounded when the agent chooses the short side, the variable that equities evaluations and risk checks already watch, an options position becomes unbounded when one leg of a structure is dropped. Its payoff of small, frequent gains and rare, large losses then looks better on a win-rate basis than the protected version. An option's value also depends on implied volatility (IV) and on the Greeks (delta, gamma, theta, vega), which the Black--Scholes model~\cite{blackscholes1973} and its descendants derive and which neither a retail user nor an LLM checks at a glance. Because volume splits across many strikes and expiries, a contract trades far less than its underlying and quotes a wider relative spread~\cite{hull2021}. Options also carry early exercise, assignment of a short leg, and pin risk at expiry, which zero-days-to-expiry contracts compress into a single session.

\rih{Five decision variables.}\label{sec:dof}
An options agent is a different target because its order carries more decisions. An equities order commits to one variable an attacker can corrupt, the direction, plus a size. An options order commits to five, namely direction (call or put, long or short), magnitude (the strike), time (the expiry), volatility (the IV paid or sold), and structure (the legs). Figure~\ref{fig:overview} shows the five variables, and Table~\ref{tab:diff} lists each difference from equities with the attack risk it adds. Rows 2--5 of the table show that every variable other than direction can carry an attack while the direction stays correct, and rows 6--8 list the market and lifecycle properties behind axes A3, A5 and A6. Each added variable is one the attacker can move while the direction stays correct.

\begin{table*}[t]
\centering
\footnotesize
\caption{Differences between options and equities, and the attack risk each difference adds. Rows 1--5 are the five decision degrees of freedom of an options order (\S\ref{sec:dof}), and rows 6--8 are properties of the options market and lifecycle. The last column points to the primitives of \S\ref{sec:primitives} and the axes of \S\ref{sec:thesis}.}
\label{tab:diff}
\setlength{\tabcolsep}{4pt}
\renewcommand{\arraystretch}{1.12}
\begin{tabularx}{\textwidth}{@{}>{\RaggedRight\arraybackslash}p{2.3cm} >{\RaggedRight\arraybackslash}p{2.9cm} >{\RaggedRight\arraybackslash}p{4.0cm} >{\RaggedRight\arraybackslash}X@{}}
\toprule
\textbf{Difference} & \textbf{Equities} & \textbf{Options} & \textbf{Added attack risk} \\
\midrule
1 Direction & Buy, sell, or hold. & Call or put, long or short, with leverage. & A framed view flips the trade, now scaled by leverage (A1). \\
2 Magnitude & Any rise is a gain. & Profit only past strike plus premium (breakeven). & An overstated catalyst sets a strike the price never reaches: right direction, still losing (O-2). \\
3 Time & No deadline; a slow thesis can still pay. & A hard expiry; time value decays faster as it nears. & A misdated catalyst ends the expiry before the event; false urgency pushes short-dated contracts (O-2, O-4). \\
4 Volatility & Price is the only input. & Premium embeds IV, which usually falls after a scheduled event. & Pre-event hype induces overpriced volatility that collapses after: right call, still losing (O-3). \\
5 Structure & One position, loss capped by size. & Multi-leg; one missing leg turns bounded loss unbounded. & Framing the protective leg as a drag on returns induces an uncovered short (O-1). \\
\midrule
6 Liquidity and price moves & Deep books, tight spreads, moves that track the stock. & Thin per-contract books, gappy quotes, prices moving in multiples of the underlying. & Market or midpoint fills pay the spread to a counterparty; small size moves the quotes the agent reads (A3, O-6). \\
7 Order-flow visibility & An agent order is anonymous in a deep book. & A lone displayed order can be the only interest in a contract; flow routes to a few wholesalers~\cite{bryzgalova2023retail}, and every print is disseminated~\cite{opra2026}. & Passive profiling of the agent's rules, then timed inputs and harvesting as its counterparty (A6, O-6). \\
8 Exercise and assignment & None. & Early exercise, assignment, and pin risk at expiry. & Induced exercise or ignored assignment converts cash or shares into an unplanned position (O-5). \\
\bottomrule
\end{tabularx}
\end{table*}

\rih{Six amplification axes.}\label{sec:thesis}
Six properties explain why the same agent error costs more, is harder to check, or is easier to exploit on options. They are leverage and non-linear payoff (A1), the Greeks and IV knowledge gap (A2), wide bid-ask spreads (A3), pricing by time decay and IV (A4), exercise, assignment, and pin (A5), and order-flow observability (A6). Table~\ref{tab:axes} pairs each with its closest equities analogue and the agent failure mode it acts on. None of them needs a new kind of agent error. Instead, the same error lands on a payoff with a lower floor (A1, A4), on a number that is harder to check (A2), or on an order that a counterparty can see (A6). A2 deserves emphasis because implied volatility is backed out of an observed price through a pricing model~\cite{blackscholes1973} and the Greeks are partial derivatives of that model. As a result, a human approver cannot check a claim such as ``IV rank is at the 80th percentile'' against a price chart, and the agent usually receives such numbers precomputed from a tool.

\begin{table*}[t]
\centering
\footnotesize
\caption{The six amplification axes. Each row pairs an options property with its closest equities analogue, states what gets strictly worse on options, and names the agent failure mode the property acts on. A1 carries the headline worst case, A2 the stealth, and A6 the reconnaissance and the counterparty channel.}
\label{tab:axes}
\renewcommand{\arraystretch}{1.15}
\begin{tabularx}{\textwidth}{@{}>{\raggedright\arraybackslash}p{2.6cm}>{\raggedright\arraybackslash}X>{\raggedright\arraybackslash}X>{\raggedright\arraybackslash}X@{}}
\toprule
\textbf{Axis} & \textbf{Equities analogue} & \textbf{Options amplification} & \textbf{Agent failure mode} \\
\midrule
\textbf{A1} Leverage and non-linear payoff &
Long stock: loss is linear and capped at the capital committed. &
One contract controls 100 shares, and a written leg without cover turns a bounded loss into one that runs to the strike (short put) or has no bound (short call). &
Treats a protective leg as a cost to trim and does not model the risk its removal re-opens. \\
\addlinespace
\textbf{A2} Greeks and IV knowledge gap &
Price, volume, and percentage moves that a supervisor can check against a chart. &
Delta, gamma, theta, vega, IV rank, expected move, and probability of profit are model-derived and cannot be checked by eye. &
Accepts a stated Greek or IV statistic without recomputing it from the chain, then sizes and structures the position on it. \\
\addlinespace
\textbf{A3} Wide bid-ask spreads &
Liquid large-cap stock trades near its quote. &
Far out-of-the-money and far-dated contracts quote wide spreads on thin open interest, so entry and exit each pay a large share of premium. &
Anchors on the midpoint, so a shifted strike or expiry costs money with the direction right. \\
\addlinespace
\textbf{A4} Pricing by time decay and IV &
Value tracks the underlying, so a correct directional call gains. &
Value also falls with time (theta) and implied volatility (vega), so a correct direction that arrives late, or after an IV collapse, still loses. &
Reasons in linear price terms and does not audit decay or a volatility change. \\
\addlinespace
\textbf{A5} Exercise, assignment, and pin &
Buy, sell, and hold, and holding is passive. &
Short legs can be assigned early, contracts near the strike carry pin risk at expiry, and exercise can discard time value. &
Mishandles involuntary events and expiry-day timing. \\
\addlinespace
\textbf{A6} Order-flow observability and illiquidity &
An agent order is anonymous in a deep, tight book. &
Per-contract books are thin, a displayed agent order may be the only visible interest, and every print is disseminated~\cite{opra2026}. &
Places predictable orders that fill against a counterparty who anticipated them. \\
\bottomrule
\end{tabularx}
\end{table*}

\rih{LLM agents at the broker.} LLM trading agents extend a language model with market-data tools, memory, and an order interface in an observe--reason--act loop over retrieved text and numbers~\cite{agentictrading2026}. On the broker side, most of them connect through MCP, which places each tool's description and return value directly in the model's context. The broker layer has opened quickly. Robinhood launched agentic trading in beta in May 2026, for equities only~\cite{robinhoodagents2026,techcrunch2026robinhood}, and later announced that agentic options trading is available to all US customers~\cite{robinhoodoptionsallcustomers2026}. However, its two official sources disagree on what those orders can be. While the support article describes the agent as placing long options orders~\cite{robinhoodsupport2026}, the order tool the agent calls documents covered calls and cash-secured puts at Level~2 and defined-risk spreads at Level~3, with uncovered writing offered at no level~\cite{robinhoodmcp2026,robinhoodlevels2026}. We score the documented tool, so Robinhood's worst case is assignment into unplanned stock, and a long-only enforcement would lower it to the premium paid.

Other brokers grant more write access. The tastytrade MCP server binds each order to a dry-run token that the same agent mints and redeems within 60 seconds, so the check confirms consistency between two calls of one agent, and uncovered writing is reachable when the account's options level allows it~\cite{tastytrademcp2026}. Alpaca's official MCP server posts option orders with no confirmation step, while Alpaca's brokerage rejects uncovered options at every options level~\cite{repo_alpacamcp,alpaca2026optionsdocs}. Thus, each broker sets an agent's worst case through two separate choices, how much the agent may write and whether anyone confirms it, and \S\ref{sec:rq2} measures both. The layer also inherits the weaknesses of MCP itself, since poisoned tool descriptions steer agent actions without touching the model~\cite{mcptox2026}. Since options agents read their chain data, Greeks, and buying power through these same tools, a poisoned field reaches the order directly.

\section{Threat Model}
\label{sec:threat}

We fix one threat model and read every system in \S\ref{sec:eval} against it. Figure~\ref{fig:overview} traces an attack from the text the attacker writes to the loss at expiry and marks the safeguards we measure at each step.

\subsection{Attacker and Victim}

\rih{Victim.} The victim is an LLM agent that reads text, calls options tools for chains, quotes, and Greeks, and holds options order authority directly or through a human who approves its tickets. Its account sits at a broker whose quote, fill, and position reports are honest. However, the agent also treats ingested text and tool-returned numbers as honest, and that assumption is what the attacker breaks.

\rih{Attacker goal.} The attacker wants the agent to send an order whose risk differs from the order it would have sent on honest inputs. We judge success by three measurable properties of the final order: (i) a defined-to-undefined conversion, in which a short option leg is left without a covering position where the pre-attack plan had bounded loss, (ii) the maximum potential loss at expiry, labeled bounded or unbounded, and (iii) the premium and spread cost lost to a mis-chosen strike, expiry, or structure. Because the direction can stay correct, a decision-flip rate misses all three.

\rih{Knowledge.} Following Kerckhoffs's principle, the attacker knows which public sources a system ingests, how its prompt, tools, and order path are designed, and every documented guardrail, because most systems in \S\ref{sec:eval} publish them. It observes public option chains, quotes, implied volatility, and prints, so it can profile an agent's triggers without access to the agent (A6). It does not see the victim's private prompt, memory, positions, or balance.

\rih{Capabilities.} The attacker writes text, and numbers embedded in text, into channels the agent reads, through the three surfaces of the FARSIGHT attack taxonomy~\cite{wang2026soktradingagentsmarket}. Through information sources (S1), it posts news, social posts, earnings commentary, unusual-options-activity alerts, and flow reports, which is indirect prompt injection~\cite{greshake2023} with a market payload. Through prompts, tools, and memory (S2), it poisons tool descriptions and tool-return fields, including returned Greeks and IV, a channel in which tool poisoning reached up to 72.8\% attack success across 20 agents~\cite{mcptox2026}. Through other agents (S3), it controls an agent whose outputs the victim acts on, such as an analyst agent that passes a volatility view to an execution agent. The content may be fabricated or accurate but selectively presented. It must read as ordinary market commentary and yield a legal order that passes whatever spending limit, margin check, or dry-run the system has. The attacker cannot tamper with the broker's reports, access the victim account, modify model weights~\cite{sleeperagents2024}, or move the underlying with its own capital, because each of these belongs to a different attacker class or to ordinary market manipulation.

\rih{Counterparty attacker.} One primitive (O-6) assumes a stronger attacker that can also post and fill orders in listed options, such as a trader with an options account or a liquidity provider. Retail options orders are mostly routed to a few wholesalers that pay for order flow~\cite{bryzgalova2023retail}, which internalize them through exchange price-improvement auctions and the limit order book~\cite{ernstspatt2026pfof,cboeaim}. As a result, a non-wholesaler reaches the agent's order when it rests on an exchange book or enters an auction, which is most likely in thin contracts. Profiling from public prints works where an agent with fixed rules dominates a contract's flow.

\subsection{Option-Only Attack Primitives}
\label{sec:primitives}

\begin{table*}[t]
\centering
\caption{Option-only attack primitives. Each primitive exploits a property of an options order that an equity order lacks. Worst-case loss is per contract (100 shares) from standard payoff mechanics~\cite{hull2021}, with $K$ the strike, $c$ the premium received, and $p$ the premium paid. Reachability distinguishes platforms that forbid uncovered writing (No-naked) from write-enabled ones such as tastytrade MCP~\cite{tastytrademcp2026}. Precursor classes refer to Table~\ref{tab:migration}.}
\label{tab:primitives}
\footnotesize
\setlength{\tabcolsep}{3.5pt}
\renewcommand{\arraystretch}{1.04}
\begin{tabularx}{\textwidth}{@{}>{\RaggedRight\arraybackslash}p{2.3cm} >{\RaggedRight\arraybackslash}X >{\centering\arraybackslash}p{0.9cm} >{\RaggedRight\arraybackslash}p{3.0cm} >{\RaggedRight\arraybackslash}p{1.9cm} >{\RaggedRight\arraybackslash}p{1.9cm} >{\centering\arraybackslash}p{1.35cm}@{}}
\toprule
\textbf{Primitive} & \textbf{Mechanism} & \textbf{Axes} & \textbf{Worst-case loss} & \textbf{No-naked} & \textbf{Write-enabled} & \textbf{Precursors} \\
\midrule
\cellcolor{paired-light-red}\textbf{O-1} Risk-structure induction (defined $\rightarrow$ undefined) & Text leads the agent to drop or skip the protective leg of a spread, or to open an uncovered short, while every stated fact stays true. & A1, A5 & Uncovered call: unbounded. Uncovered put: $(K-c)\times 100$. & \xmark\ Uncovered legs forbidden. & \cmark\ Reachable. & M2, M3, M6, M7, M8 \\
\textbf{O-2} Strike and expiry displacement & The direction is kept, but the choice moves to a farther out-of-the-money strike, a nearer expiry, or a thinner contract. & A3, A4 & Long: $p\times 100$, with a higher chance of losing all of it. Short: strike moves toward the money. & \cmark\ Reachable; premium or assignment. & \cmark\ Reachable. & M1, M4, M5 \\
\textbf{O-3} Greeks and volatility misstatement & IV rank, expected move, delta, or breakeven are misstated or re-presented, so the agent buys expensive volatility, sells cheap volatility, or sizes on a wrong delta. & A2, A4 & Long: overpaid premium. Short: position sized on understated risk, up to the O-1 bound. & \cmark\ Reachable; premium or assignment. & \cmark\ Reachable. & M2, M4, M5, M6, M8 \\
\textbf{O-4} Expiry-pressure triggering & Urgency or complacency about days to expiry and time decay leads the agent to chase, roll at a loss, or hold a decaying position to expiry. & A4 & Long: $p\times 100$ lost to decay. Short: held into an adverse move near expiry. & \cmark\ Reachable; premium or assignment. & \cmark\ Reachable. & M1, M3, M9 \\
\textbf{O-5} Exercise, assignment, and pin exploitation & The agent is induced to exercise early and forfeit time value, to ignore early-assignment risk on a short leg, or to mishandle a position pinned near the strike at expiry. & A5, A1 & Long: forfeited time value. Short: assignment leaves an unplanned share position of 100 per contract. & \cmark\ Reachable via covered and cash-secured legs. & \cmark\ Reachable. & M3, M7, M9 \\
\cellcolor{paired-light-orange}\textbf{O-6} Profile and harvest & The attacker reads the agent's displayed orders and prints, infers its triggers, times an input to induce an order in a thin contract, and fills it on the other side. & A6, A3 & Spread plus induced mispricing per fill, repeatable. Long: up to $p\times 100$. & \cmark\ Reachable (no short leg needed). & \cmark\ Reachable. & M1, M4, M5 \\
\bottomrule
\end{tabularx}
\end{table*}

\begin{table*}[t]
\centering
\caption{Migration map from equities to options. Each row is a publicly documented attack class, grouped by the attack taxonomy of the equities SoK~\cite{wang2026soktradingagentsmarket}. The effect column reports what the exemplar measured on its own victims (equities trading agents, pricing agents, or general LLM agents); no number in this table was measured on an options agent. The escalation column follows from options payoff mechanics (\S\ref{sec:thesis}). Axes are those of \S\ref{sec:thesis}, and primitives follow Table~\ref{tab:primitives}.}
\label{tab:migration}
\footnotesize
\setlength{\tabcolsep}{3.5pt}
\renewcommand{\arraystretch}{1.04}
\begin{tabularx}{\textwidth}{@{}>{\RaggedRight\arraybackslash}p{2.35cm} >{\RaggedRight\arraybackslash}p{1.7cm} >{\RaggedRight\arraybackslash}X >{\RaggedRight\arraybackslash}X >{\centering\arraybackslash}p{0.95cm} >{\centering\arraybackslash}p{1.0cm}@{}}
\toprule
\textbf{Attack class} & \textbf{Public exemplar} & \textbf{Documented effect (victim type)} & \textbf{Options escalation} & \textbf{Axes} & \textbf{Feeds} \\
\midrule
\multicolumn{6}{@{}l}{\cellcolor{paired-light-blue}\textit{Attacks on information sources}} \\
M1 Fabricated news and content injection & TradeTrap~\cite{tradetrap2025} & Counterfeit news cut an adaptive equities agent's return from 11.59\% to 5.26\% and its Sharpe ratio from 4.34 to 3.20 (equities agent). & A fabricated catalyst sets the expected move, so a long option bought ahead of it loses to time decay even if the underlying stays flat. & A1, A4 & O-2, O-4 \\
M2 Retrieval and corpus poisoning & PoisonedRAG \cite{zou2025poisonedrag} & Five injected texts per target question reached 90\% attack success over millions of documents (general RAG). & A poisoned playbook entry can present an uncovered structure as the standard form of a defined-risk strategy, which the agent cannot check (A2). & A1, A2 & O-1, O-3 \\
M3 Misleading-but-true framing & Framing and bias studies~\cite{tversky1981framing,itzhak2024instructed} & Instruction-tuned models show stronger decoy, certainty, and belief biases than their base models (general LLMs). & Short-premium payoffs win often and lose rarely but large, a profile certainty-effect framing favors, pointing the agent toward writing options. & A1, A4 & O-1, O-4 \\
M4 Presentation and market-signal injection & Market Signal Injection~\cite{marketsignalinjection2026} & Changing number formatting, competitor ordering, or commentary, with data values unchanged, shifted pricing decisions across 12 models (pricing agent). & The manipulated quantities become IV rank, expected move, and Greeks, whose magnitude neither a supervisor nor the model can confirm by inspection. & A2, A4 & O-3 \\
M5 Manipulated provenance and manufactured consensus & Adversarial SEO~\cite{nestaas2025adversarialseo}, source attribution~\cite{germani2025sourceframing}, social bots~\cite{ferrara2016rise} & Crafted content steered production LLM search engines, and re-attributing the same text to another source shifted LLM judgments (general LLMs). & Options-flow and ``unusual activity'' reports aggregate anonymous prints with weak provenance, so a consensus volatility narrative is cheap to manufacture. & A2, A3 & O-2, O-3 \\
M6 Visual-channel manipulation & Self-interpreting images~\cite{selfinterpreting2025} & Images with hidden meta-instructions steered a VLM's answers while they stayed plausible (general VLMs). & Payoff diagrams, volatility surfaces, and chain screenshots are image inputs, and a misread chart hides the tail of a written leg. & A1, A2 & O-1, O-3 \\
\midrule
\multicolumn{6}{@{}l}{\cellcolor{paired-light-orange}\textit{Attacks on agents}} \\
M7 Indirect and hidden-content prompt injection & Greshake et al.~\cite{greshake2023} & Instructions planted in retrieved content controlled API calls of deployed LLM applications, including Bing Chat (general agents). & An options ticket has more fields (legs, ratio, strike, expiry, open or close, exercise), so an injected ticket can alter one and leave the rest plausible. & A1, A5 & O-1, O-5 \\
M8 Tool-response, metadata, and state poisoning & MCPTox~\cite{mcptox2026}, TradeTrap~\cite{tradetrap2025} & Poisoned tool metadata reached up to 72.8\% attack success across 20 agents (general MCP agents). Tampered state feedback turned a procedural equities agent's 0.91\% return into a 61.02\% loss, with position utilization rising from 20.78\% to 646.86\% (equities agent). & Options tools return Greeks, buying power, margin requirement, and maximum loss, so a falsified risk field disables the one check that bounds exposure before the order leaves. & A1, A2 & O-1, O-3 \\
M9 Persistent backdoor and sleeper behavior & Sleeper Agents~\cite{sleeperagents2024}, agent backdoors~\cite{yang2024watchout} & Backdoors persisted through safety training, and triggers hidden in environment observations evaded textual backdoor defenses (general LLMs and agents). & Triggers can key on options-only state such as days to expiry, which an equities-style backtest never holds, so the trigger fires when gamma and theta peak. & A4, A5 & O-4, O-5 \\
\bottomrule
\end{tabularx}
\end{table*}

An options order lets the attacker corrupt four of the five decision variables while the direction stays correct. The agent buys the call it meant to buy, but at a strike the move never reaches, on an expiry that ends before the catalyst, at an implied volatility that collapses after the event, or as a spread without the leg that bounded its loss. Table~\ref{tab:primitives} turns these variables into six primitives. Five pair a variable with the payoff mechanism that turns a corrupted value into loss, and the sixth targets the observability of the agent's own orders. Nine publicly documented attack classes against equities, pricing, and general-purpose agents reach these primitives through the surfaces above (Table~\ref{tab:migration}), and in each one the options payoff turns the same corruption into a larger or harder-to-detect loss. Appendix~\ref{app:migration} groups these escalations into three patterns.

\rih{O-1: risk-structure induction.} O-1 is the worst case we consider, the one corruption whose loss has no ceiling while the agent's direction stays correct. Consider a bull put spread that sells the 100-strike put and buys the 95-strike put for a net credit of \$1.00 per share, so its maximum loss is $(5-1)\times 100=\$400$ per contract (Figure~\ref{fig:payoff}a). If the long 95 put is removed, the position becomes an uncovered put with a maximum loss of $(100-1)\times 100=\$9{,}900$. Removing the protective call from a bear call spread leaves no upper bound at all. The inducing text needs no falsehood, because presenting the protective leg as a cost that rarely pays out is true in most expiries, and an injected ticket or a falsified maximum-loss field reaches the same end state without persuading the model.

\rih{O-2 to O-5: loss with the direction right.} Strike and expiry displacement (O-2) moves the agent to a farther out-of-the-money strike, a nearer expiry, or a thinner contract, so the position expires worthless more often while it still matches the agent's own analysis. Greeks and volatility misstatement (O-3) targets the numbers that decide whether an option is cheap, which few supervisors can check (A2). As a result, the agent overpays for volatility that collapses after the event (Figure~\ref{fig:payoff}b) or sizes a short position on an understated delta. Expiry-pressure triggering (O-4) uses urgency or complacency about time decay, which speeds up near expiry~\cite{hull2021}, to make the agent chase, roll at a loss, or hold a decaying position. Exercise and assignment exploitation (O-5) induces an early exercise that forfeits time value, or leads the agent to ignore early-assignment and pin risk on a short leg, which can leave an unplanned position of 100 shares per contract.

\rih{O-6: profile and harvest.} O-6 pays the attacker directly and needs no uncovered leg. When a fixed-rule agent dominates a thin contract's flow, its displayed orders and the consolidated prints~\cite{opra2026} reveal its entry conditions, strikes, and timing (A6). The counterparty attacker times an input from any precursor class so that the agent enters the contract, and then fills the order on the other side, collecting the spread and the induced mispricing. Because each trade is small and ordinary, the attack repeats.

\rih{Reachability.} The broker's write policy decides which primitives an agent can reach. On a platform that forbids uncovered writing, such as Robinhood's documented order tool or Alpaca (\S\ref{sec:background}), O-1 is out of reach, O-2 to O-4 are bounded by the premium or by the assignment exposure of a covered leg, and O-5 and O-6 stay reachable. By contrast, all six are reachable on tastytrade's MCP server~\cite{tastytrademcp2026}. Hence, a ban on uncovered writing removes one primitive and leaves five, which is why \S\ref{sec:rq2} measures write scope and the worst reachable loss for every system.

\section{Measurement Methodology}
\label{sec:method}

\subsection{Data Collection}

\rih{Sources and snapshot.} We collected candidate systems from three sources at one snapshot, and every count in this paper refers to it. On GitHub, six repository searches through the search API (strings in Appendix~\ref{app:search}) returned \nQueryRaw{} hits and \nUnion{} unique repositories, each query staying under the API's 1,000-result cap. For brokers, we collected first-party agent interfaces, seeded by an industry test that found agent MCP servers at 7 of 14 brokers~\cite{stockbrokers2026}, and read each broker's documentation. For the research literature, we searched arXiv and security, machine-learning, and finance proceedings with options terms combined with LLM and agent terms and followed references from recent trading-agent work~\cite{wang2026soktradingagentsmarket,agentictrading2026}.

\rih{Eligibility.} A candidate is eligible if (I1) an LLM makes, proposes, or gates an options decision (strategy, strike, expiry, structure, entry, or exit), or an interface is published for an LLM agent to do so, (I2) it reads option-chain or option-position data, (I3) a public paper, README, or documentation page supports scoring it, and (I4) the artifact is available on the snapshot date.

\rih{Screening.} Figure~\ref{fig:funnel} traces the GitHub screen. First, a keyword filter keeps a repository if its name, description, topics, and README match at least one of 21 LLM terms and one of 29 options terms, which leaves \nStageOneKept{} of the \nUnion{} repositories. Because the filter favors recall, we then read every survivor against I1--I4 and coded each exclusion with one of seven reasons in a per-repository screening log (Appendix~\ref{app:search}). The reading pass leaves \nEligible{} eligible repositories. The most common exclusions are software with no trading system (\nExcludeEfive{}), options bots with no LLM in the decision path (\nExcludeEtwo{}), and duplicates, forks, or mirrors of an eligible project (\nExcludeEsix{}).

\begin{figure}[!htb]
\centering
\begin{tikzpicture}[font=\sffamily\small]
\fill[okblue!35] (0,0.00) rectangle (3.90,-0.36);
\node[anchor=west, inner sep=2pt] at (3.90,-0.18) {\bfseries \nQueryRaw{}};
\node[anchor=east, inner sep=3pt] at (0,-0.18) {query hits, six strings};
\fill[okblue!45] (0,-0.52) rectangle (2.42,-0.88);
\node[anchor=west, inner sep=2pt] at (2.42,-0.70) {\bfseries \nUnion{}};
\node[anchor=east, inner sep=3pt] at (0,-0.70) {unique repositories};
\fill[okblue!55] (0,-1.04) rectangle (1.48,-1.40);
\node[anchor=west, inner sep=2pt] at (1.48,-1.22) {\bfseries \nStageOneKept{}};
\node[anchor=east, inner sep=3pt] at (0,-1.22) {pass the keyword filter};
\fill[okblue!65] (0,-1.56) rectangle (0.68,-1.92);
\node[anchor=west, inner sep=2pt] at (0.68,-1.74) {\bfseries \nEligible{}};
\node[anchor=east, inner sep=3pt] at (0,-1.74) {eligible after reading (I1--I4)};
\fill[okverm!70] (0,-2.08) rectangle (0.12,-2.44);
\node[anchor=west, inner sep=2pt] at (0.12,-2.26) {\bfseries 14};
\node[anchor=east, inner sep=3pt] at (0,-2.26) {scored from the union};
\node[anchor=north west, inner sep=0pt, text width=7.9cm, align=left, font=\sffamily\small, execute at begin node={\hyphenpenalty=10000\exhyphenpenalty=10000}] at (-3.55,-2.65) {The other 12 of the \nSchemes{} scored systems come from broker \\ documentation, searches outside the six strings, and the academic search.};
\end{tikzpicture}
\caption{\textbf{GitHub screening funnel.} Counts from the logged screen of the six-query union. Bars are proportional to the number of repositories at each stage, and \nEligibleUnscored{} eligible repositories remain unscored.}
\label{fig:funnel}
\end{figure}

\rih{Ecosystem metadata.} For RQ1 we record each eligible repository's creation date, star count, and language from the GitHub API, the LLM and options terms that the keyword filter matched in its name, description, topics, or README, and whether its own name, description, or topics mention a hackathon or a broker. We count what a repository says about itself, so every keyword count can be recomputed from the logged metadata. Since creation dates describe the repositories that still existed at the snapshot, a deleted repository leaves no trace in them.

\rih{Scored sample.} Because scoring all \nEligible{} repositories in depth is infeasible, we score a stratified sample of \nSchemes{} systems. Three strata are censuses of what we found: every broker-native agent interface (\nBroker{}), every academic system meeting I1--I4 (\nAcademic{}), and every eligible open-source project with at least \starThreshold{} stars (\nOSSAdopted{}). The threshold falls on a natural break, since the next eligible repository below it has \nextStarBelow{} stars. The fourth stratum is purposive. Its \nOSSPurposive{} lower-star repositories cover design points that the censuses lack, such as veto-only autonomy, paper-only execution, advisory-only agents, and crypto or index options. Therefore, we report prevalence from the \nCensus{} census systems or from all \nSchemes{}, and use the purposive stratum to describe designs. Of the \nExamined{} candidates examined in depth, \nExcludedReview{} were excluded after review (Appendix~\ref{app:schemes}).

\subsection{The Safeguard Codebook}
\label{sec:gamma}

\begin{figure*}[!t]
\centering
\includegraphics[width=\textwidth]{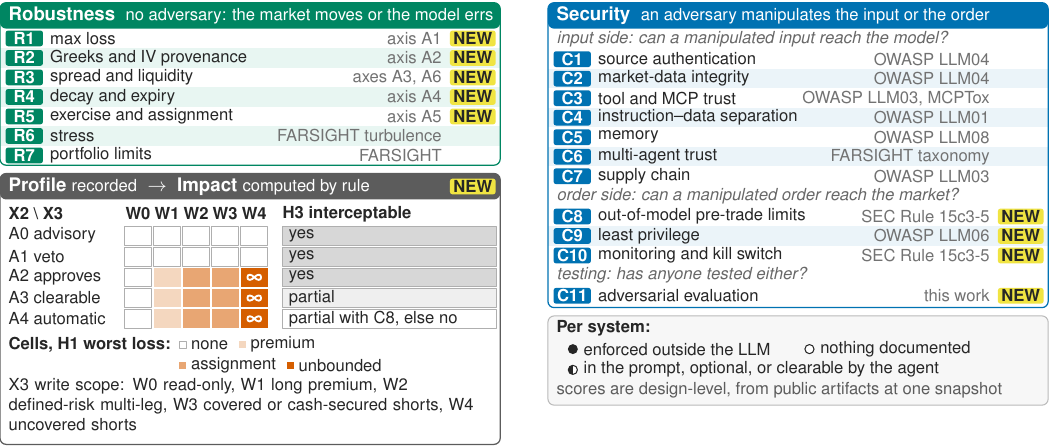}
\caption{\textbf{The safeguard codebook.} Seven robustness indicators (R) ask whether a system stops a harmful order when the market moves or the model errs, and eleven security indicators (C) ask whether a manipulated input or order can reach the model or the market. Each system gets one mark per indicator: filled if the safeguard is enforced outside the LLM, half-filled if it lives in the prompt, is optional, or can be cleared by the agent, and empty if nothing is documented. The grid at the bottom left shows how autonomy (X2, rows) and write scope (X3, columns) fix the worst reachable loss (H1, cell shading) and whether an order can be intercepted (H3, right of the grid). NEW marks indicators with no counterpart among the metrics FARSIGHT scores.}
\label{fig:codebook}
\end{figure*}

We score each system with a design-level codebook built on the two dimensions of FARSIGHT~\cite{wang2026soktradingagentsmarket}, robustness and security, and redefined for the options order. As Figure~\ref{fig:codebook} shows, the codebook asks two questions of every system, whether it stops a harmful order when nothing adversarial happens and whether it stops one that an attacker induced, and then records what the system can reach. Table~\ref{tab:codebook-full} in Appendix~\ref{app:codebook} defines every indicator together with the public framework it operationalizes, namely the OWASP Top 10 for LLM Applications~\cite{owasp2025llm}, the market-access Rule 15c3-5 of the U.S. Securities and Exchange Commission (SEC)~\cite{sec2010rule15c3_5}, and Rule 2360, under which FINRA, the U.S. self-regulator of broker-dealers, approves customers for options~\cite{finra2360}.

\rih{Robustness (R1--R7).} Seven indicators ask whether a system stops the harmful orders that an honest agent can produce on options when the market moves or the model errs. R1--R5 follow the axes A1--A5, from a maximum loss computed before entry (R1) and Greeks and IV derived from a pricing model (R2) to rules for exercise and assignment (R5). R6 (stress) and R7 (portfolio limits) come from FARSIGHT. The same checks also stop a manipulated order, whatever produced it.

\rih{Security (C1--C11).} C1--C7 guard the inputs that arrive through the three surfaces of \S\ref{sec:threat}, covering source authentication, market-data integrity, tool trust, instruction--data separation, memory, peer agents, and supply chain. C8--C10 guard the order with a pre-trade limit outside the model, least privilege, and monitoring with a kill switch. C11 records whether anyone tested either side adversarially.

\rih{Profile and impact.} Five profile attributes record what a system can reach without judging it. Autonomy (X2) runs from A0 (advisory) to A4 (no gate), where A3 is a gate that the agent can clear, that is off by default, or that a documented setting removes. Write scope (X3) runs from W0 (read-only) through W1 (long premium), W2 (defined-risk spreads), and W3 (covered or cash-secured shorts) to W4 (uncovered shorts), counting the broker's own restrictions. Three impact indicators are computed from the profile, namely the worst loss the order path can reach (H1), the blast radius (H2), and whether a control outside the model can stop a manipulated order before execution (H3).

\rih{Scoring levels.} Each robustness and security indicator takes one of three levels. \fullc{} means that the safeguard is documented and enforced outside the LLM, in code, configuration, or platform policy. \halfc{} means that it is documented but depends on the model following an instruction, is optional or off by default, or can be cleared by the agent itself. \emptyc{} means that the public artifacts document nothing, and C8 and C10 do not apply to a system without an order path. The line between full and partial is the one that matters for security, because an instruction in a prompt sits in the same context the attacker writes to.

\rih{Scoring process and validation.} One coder scored the systems of an initial snapshot from their papers, repositories, and documentation, and the \nAddedSchemes{} systems added at the final snapshot were scored with LLM assistance from the same public artifacts. Every cell cites its evidence with a quotation and a source locator (Appendix~\ref{app:schemes}). Five decision rules make the profile and impact cells mechanical (Appendix~\ref{app:codebook}), and a consistency check rejects any cell that violates them. After scoring, a second LLM-assisted pass rechecked the autonomy, write-scope, and pre-trade-limit cells of the added systems against their code and documentation, and every cell it changed keeps its previous value in the evidence file. The scores describe documented designs. As a result, an undocumented safeguard in a running deployment scores \emptyc{}, and a documented safeguard that a deployment has switched off still counts as present.

\section{Findings}
\label{sec:eval}

\begin{table*}[tp]
\centering
\caption{Safeguard scores for the \nSchemes{} included options-agent systems (at the snapshot), grouped by sampling stratum. The academic and broker-native groups and the open-source group with at least \starThreshold{} GitHub stars are censuses, and the lower-star open-source group is a purposive sample that supports no prevalence claim. Safeguard cells: \capbox{okblue!62}{\textcolor{white}{\fullc}}~enforced outside the model, \capbox{oksky!32}{\halfc}~model-dependent, optional, or clearable by the agent, \emptyc{}~not documented, \capbox{black!9}{--}~not applicable (no order path). Shading marks automatic autonomy (A3, A4), uncovered write scope (W4), and the worst reachable loss: \capbox{okverm!80}{\textcolor{white}{unb.}}~unbounded, \capbox{okorange!45}{asgn.}~assignment, \capbox{okyellow!55}{prem.}~premium. H1--H3 follow the rules of Appendix~\ref{app:codebook}, and H3 uses the safeguard colors. $^\dagger$Hosted service whose repository publishes the interface, not the server. Per-cell evidence is in Appendix~\ref{app:schemes}.}
\label{tab:schemes}
\footnotesize
\setlength{\tabcolsep}{2.05pt}
\renewcommand{\arraystretch}{1.05}
\begin{tabular}{@{}rlccccccccccccccccccccccc@{}}
\toprule
& & \multicolumn{2}{c}{\textbf{Profile}} & \multicolumn{3}{c}{\textbf{Impact}} & \multicolumn{7}{c}{\textbf{Robustness}} & \multicolumn{11}{c}{\textbf{Security}} \\
\cmidrule(lr){3-4}\cmidrule(lr){5-7}\cmidrule(lr){8-14}\cmidrule(l){15-25}
\# & \textbf{System} & X2 & X3 & H1 & H2 & H3 & R1 & R2 & R3 & R4 & R5 & R6 & R7 & C1 & C2 & C3 & C4 & C5 & C6 & C7 & C8 & C9 & C10 & C11 \\
\midrule
\multicolumn{25}{@{}l}{\textit{Academic}} \\
1 & {\sffamily\fontsize{8.9}{9.5}\selectfont Kuang and Lin}~\cite{kuanglin2025wheel} & A0 & W0 & none & user & \cellcolor{okblue!62}\textcolor{white}{\fullc} & \emptyc & \emptyc & \cellcolor{oksky!32}\halfc & \cellcolor{oksky!32}\halfc & \cellcolor{oksky!32}\halfc & \cellcolor{oksky!32}\halfc & \cellcolor{oksky!32}\halfc & \emptyc & \emptyc & \emptyc & \emptyc & \cellcolor{oksky!32}\halfc & \cellcolor{oksky!32}\halfc & \cellcolor{oksky!32}\halfc & \cellcolor{black!9}\textcolor{black!55}{--} & \emptyc & \cellcolor{black!9}\textcolor{black!55}{--} & \emptyc \\
2 & {\sffamily\fontsize{8.9}{9.5}\selectfont OQL}~\cite{oql2026} & A0 & W0 & none & user & \cellcolor{okblue!62}\textcolor{white}{\fullc} & \cellcolor{oksky!32}\halfc & \cellcolor{oksky!32}\halfc & \emptyc & \cellcolor{oksky!32}\halfc & \emptyc & \emptyc & \emptyc & \emptyc & \emptyc & \emptyc & \emptyc & \emptyc & \emptyc & \emptyc & \cellcolor{black!9}\textcolor{black!55}{--} & \emptyc & \cellcolor{black!9}\textcolor{black!55}{--} & \emptyc \\
\multicolumn{25}{@{}l}{\textit{Broker-native}} \\
3 & {\sffamily\fontsize{8.9}{9.5}\selectfont OKX Agent Trade Kit}~\cite{repo_okxagenttradekit} & \cellcolor{okverm!16}A4 & \cellcolor{okverm!16}W4 & \cellcolor{okverm!80}\textcolor{white}{\textbf{unb.}} & plat. & \emptyc & \emptyc & \cellcolor{oksky!32}\halfc & \emptyc & \emptyc & \emptyc & \emptyc & \emptyc & \emptyc & \emptyc & \cellcolor{oksky!32}\halfc & \emptyc & \emptyc & \emptyc & \cellcolor{oksky!32}\halfc & \emptyc & \cellcolor{oksky!32}\halfc & \cellcolor{oksky!32}\halfc & \emptyc \\
4 & {\sffamily\fontsize{8.9}{9.5}\selectfont Public MCP}~\cite{repo_publicmcp} & \cellcolor{okverm!16}A4 & \cellcolor{okverm!16}W4 & \cellcolor{okverm!80}\textcolor{white}{\textbf{unb.}} & plat. & \emptyc & \emptyc & \cellcolor{oksky!32}\halfc & \cellcolor{oksky!32}\halfc & \emptyc & \emptyc & \emptyc & \emptyc & \emptyc & \emptyc & \cellcolor{oksky!32}\halfc & \emptyc & \emptyc & \emptyc & \cellcolor{oksky!32}\halfc & \emptyc & \cellcolor{oksky!32}\halfc & \cellcolor{oksky!32}\halfc & \emptyc \\
5 & {\sffamily\fontsize{8.9}{9.5}\selectfont Webull MCP}~\cite{repo_webullmcp} & \cellcolor{okverm!16}A4 & \cellcolor{okverm!16}W4 & \cellcolor{okverm!80}\textcolor{white}{\textbf{unb.}} & plat. & \emptyc & \emptyc & \cellcolor{oksky!32}\halfc & \emptyc & \emptyc & \emptyc & \emptyc & \emptyc & \emptyc & \emptyc & \cellcolor{oksky!32}\halfc & \emptyc & \emptyc & \emptyc & \cellcolor{oksky!32}\halfc & \emptyc & \cellcolor{oksky!32}\halfc & \cellcolor{oksky!32}\halfc & \emptyc \\
6 & {\sffamily\fontsize{8.9}{9.5}\selectfont tastytrade MCP}~\cite{tastytrademcp2026} & \cellcolor{okverm!16}A3 & \cellcolor{okverm!16}W4 & \cellcolor{okverm!80}\textcolor{white}{\textbf{unb.}} & plat. & \cellcolor{oksky!32}\halfc & \cellcolor{oksky!32}\halfc & \cellcolor{oksky!32}\halfc & \cellcolor{oksky!32}\halfc & \cellcolor{oksky!32}\halfc & \cellcolor{oksky!32}\halfc & \cellcolor{oksky!32}\halfc & \cellcolor{oksky!32}\halfc & \cellcolor{oksky!32}\halfc & \emptyc & \cellcolor{okblue!62}\textcolor{white}{\fullc} & \cellcolor{oksky!32}\halfc & \cellcolor{oksky!32}\halfc & \emptyc & \cellcolor{okblue!62}\textcolor{white}{\fullc} & \cellcolor{okblue!62}\textcolor{white}{\fullc} & \cellcolor{oksky!32}\halfc & \cellcolor{oksky!32}\halfc & \cellcolor{oksky!32}\halfc \\
7 & {\sffamily\fontsize{8.9}{9.5}\selectfont IBKR AI integration}~\cite{ibkraiintegration2026} & A2 & \cellcolor{okverm!16}W4 & \cellcolor{okverm!80}\textcolor{white}{\textbf{unb.}} & plat. & \cellcolor{okblue!62}\textcolor{white}{\fullc} & \emptyc & \emptyc & \emptyc & \emptyc & \emptyc & \emptyc & \emptyc & \emptyc & \emptyc & \emptyc & \emptyc & \emptyc & \emptyc & \emptyc & \cellcolor{okblue!62}\textcolor{white}{\fullc} & \cellcolor{oksky!32}\halfc & \emptyc & \emptyc \\
8 & {\sffamily\fontsize{8.9}{9.5}\selectfont TradeStation MCP}~\cite{tradestationmcp2026} & A2 & \cellcolor{okverm!16}W4 & \cellcolor{okverm!80}\textcolor{white}{\textbf{unb.}} & plat. & \cellcolor{okblue!62}\textcolor{white}{\fullc} & \emptyc & \emptyc & \emptyc & \emptyc & \emptyc & \emptyc & \emptyc & \emptyc & \emptyc & \emptyc & \emptyc & \emptyc & \emptyc & \emptyc & \cellcolor{okblue!62}\textcolor{white}{\fullc} & \cellcolor{oksky!32}\halfc & \cellcolor{oksky!32}\halfc & \emptyc \\
9 & {\sffamily\fontsize{8.9}{9.5}\selectfont Alpaca MCP}~\cite{repo_alpacamcp} & \cellcolor{okverm!16}A4 & W3 & \cellcolor{okorange!45}asgn. & plat. & \cellcolor{oksky!32}\halfc & \cellcolor{okblue!62}\textcolor{white}{\fullc} & \cellcolor{oksky!32}\halfc & \cellcolor{oksky!32}\halfc & \emptyc & \cellcolor{oksky!32}\halfc & \emptyc & \emptyc & \cellcolor{oksky!32}\halfc & \emptyc & \cellcolor{oksky!32}\halfc & \cellcolor{oksky!32}\halfc & \emptyc & \emptyc & \cellcolor{oksky!32}\halfc & \cellcolor{oksky!32}\halfc & \cellcolor{oksky!32}\halfc & \emptyc & \cellcolor{oksky!32}\halfc \\
10 & {\sffamily\fontsize{8.9}{9.5}\selectfont Robinhood agentic options}~\cite{robinhoodoptions2026,robinhoodmcp2026} & \cellcolor{okverm!16}A3 & W3 & \cellcolor{okorange!45}asgn. & plat. & \cellcolor{oksky!32}\halfc & \cellcolor{okblue!62}\textcolor{white}{\fullc} & \cellcolor{oksky!32}\halfc & \cellcolor{oksky!32}\halfc & \cellcolor{oksky!32}\halfc & \cellcolor{oksky!32}\halfc & \cellcolor{oksky!32}\halfc & \cellcolor{oksky!32}\halfc & \emptyc & \emptyc & \emptyc & \emptyc & \emptyc & \emptyc & \emptyc & \cellcolor{oksky!32}\halfc & \cellcolor{okblue!62}\textcolor{white}{\fullc} & \cellcolor{okblue!62}\textcolor{white}{\fullc} & \emptyc \\
\multicolumn{25}{@{}l}{\textit{Open source, at least \starThreshold{} stars}} \\
11 & {\sffamily\fontsize{8.9}{9.5}\selectfont lumibot}~\cite{repo_lumibot} & \cellcolor{okverm!16}A4 & \cellcolor{okverm!16}W4 & \cellcolor{okverm!80}\textcolor{white}{\textbf{unb.}} & plat. & \cellcolor{oksky!32}\halfc & \cellcolor{oksky!32}\halfc & \cellcolor{oksky!32}\halfc & \cellcolor{oksky!32}\halfc & \cellcolor{oksky!32}\halfc & \cellcolor{oksky!32}\halfc & \cellcolor{oksky!32}\halfc & \cellcolor{oksky!32}\halfc & \cellcolor{oksky!32}\halfc & \emptyc & \cellcolor{oksky!32}\halfc & \cellcolor{oksky!32}\halfc & \cellcolor{oksky!32}\halfc & \cellcolor{oksky!32}\halfc & \cellcolor{oksky!32}\halfc & \cellcolor{oksky!32}\halfc & \cellcolor{oksky!32}\halfc & \cellcolor{oksky!32}\halfc & \cellcolor{oksky!32}\halfc \\
12 & {\sffamily\fontsize{8.9}{9.5}\selectfont tasty-agent}~\cite{repo_tastyagent} & \cellcolor{okverm!16}A4 & \cellcolor{okverm!16}W4 & \cellcolor{okverm!80}\textcolor{white}{\textbf{unb.}} & plat. & \emptyc & \cellcolor{oksky!32}\halfc & \cellcolor{oksky!32}\halfc & \cellcolor{oksky!32}\halfc & \cellcolor{oksky!32}\halfc & \emptyc & \cellcolor{oksky!32}\halfc & \emptyc & \emptyc & \emptyc & \cellcolor{oksky!32}\halfc & \emptyc & \emptyc & \emptyc & \cellcolor{oksky!32}\halfc & \emptyc & \emptyc & \cellcolor{oksky!32}\halfc & \emptyc \\
13 & {\sffamily\fontsize{8.9}{9.5}\selectfont openalgo}~\cite{repo_openalgo} & \cellcolor{okverm!16}A3 & \cellcolor{okverm!16}W4 & \cellcolor{okverm!80}\textcolor{white}{\textbf{unb.}} & plat. & \cellcolor{oksky!32}\halfc & \cellcolor{oksky!32}\halfc & \cellcolor{okblue!62}\textcolor{white}{\fullc} & \cellcolor{oksky!32}\halfc & \cellcolor{oksky!32}\halfc & \emptyc & \cellcolor{oksky!32}\halfc & \cellcolor{oksky!32}\halfc & \cellcolor{oksky!32}\halfc & \emptyc & \cellcolor{oksky!32}\halfc & \cellcolor{oksky!32}\halfc & \cellcolor{oksky!32}\halfc & \emptyc & \cellcolor{oksky!32}\halfc & \cellcolor{oksky!32}\halfc & \cellcolor{oksky!32}\halfc & \cellcolor{okblue!62}\textcolor{white}{\fullc} & \emptyc \\
14 & {\sffamily\fontsize{8.9}{9.5}\selectfont schwab-mcp}~\cite{repo_schwabmcp} & \cellcolor{okverm!16}A3 & \cellcolor{okverm!16}W4 & \cellcolor{okverm!80}\textcolor{white}{\textbf{unb.}} & plat. & \cellcolor{oksky!32}\halfc & \emptyc & \cellcolor{oksky!32}\halfc & \emptyc & \emptyc & \emptyc & \emptyc & \emptyc & \emptyc & \emptyc & \emptyc & \emptyc & \emptyc & \emptyc & \cellcolor{oksky!32}\halfc & \cellcolor{oksky!32}\halfc & \cellcolor{oksky!32}\halfc & \cellcolor{oksky!32}\halfc & \emptyc \\
15 & {\sffamily\fontsize{8.9}{9.5}\selectfont trade-it-mcp}$^\dagger$~\cite{repo_tradeitmcp} & \cellcolor{okverm!16}A3 & \cellcolor{okverm!16}W4 & \cellcolor{okverm!80}\textcolor{white}{\textbf{unb.}} & plat. & \cellcolor{oksky!32}\halfc & \emptyc & \emptyc & \emptyc & \emptyc & \emptyc & \emptyc & \emptyc & \emptyc & \emptyc & \emptyc & \emptyc & \emptyc & \emptyc & \emptyc & \cellcolor{oksky!32}\halfc & \cellcolor{oksky!32}\halfc & \cellcolor{oksky!32}\halfc & \emptyc \\
16 & {\sffamily\fontsize{8.9}{9.5}\selectfont OpenProphet}~\cite{repo_openprophet} & \cellcolor{okverm!16}A4 & W3 & \cellcolor{okorange!45}asgn. & plat. & \cellcolor{oksky!32}\halfc & \emptyc & \cellcolor{oksky!32}\halfc & \cellcolor{oksky!32}\halfc & \cellcolor{oksky!32}\halfc & \emptyc & \cellcolor{oksky!32}\halfc & \cellcolor{oksky!32}\halfc & \emptyc & \emptyc & \cellcolor{oksky!32}\halfc & \emptyc & \emptyc & \emptyc & \emptyc & \cellcolor{oksky!32}\halfc & \emptyc & \cellcolor{oksky!32}\halfc & \emptyc \\
17 & {\sffamily\fontsize{8.9}{9.5}\selectfont trading\_skills}~\cite{repo_tradingskills} & \cellcolor{okverm!16}A3 & W2 & \cellcolor{okorange!45}asgn. & plat. & \cellcolor{oksky!32}\halfc & \cellcolor{oksky!32}\halfc & \cellcolor{oksky!32}\halfc & \cellcolor{okblue!62}\textcolor{white}{\fullc} & \cellcolor{oksky!32}\halfc & \cellcolor{oksky!32}\halfc & \cellcolor{oksky!32}\halfc & \cellcolor{oksky!32}\halfc & \emptyc & \emptyc & \emptyc & \emptyc & \emptyc & \emptyc & \cellcolor{oksky!32}\halfc & \cellcolor{okblue!62}\textcolor{white}{\fullc} & \cellcolor{oksky!32}\halfc & \cellcolor{oksky!32}\halfc & \emptyc \\
\multicolumn{25}{@{}l}{\textit{Open source, fewer than \starThreshold{} stars (purposive sample)}} \\
18 & {\sffamily\fontsize{8.9}{9.5}\selectfont TastyScanner MCP}~\cite{repo_tastyscanner} & \cellcolor{okverm!16}A4 & \cellcolor{okverm!16}W4 & \cellcolor{okverm!80}\textcolor{white}{\textbf{unb.}} & user & \emptyc & \cellcolor{oksky!32}\halfc & \cellcolor{oksky!32}\halfc & \cellcolor{oksky!32}\halfc & \cellcolor{oksky!32}\halfc & \emptyc & \cellcolor{oksky!32}\halfc & \emptyc & \emptyc & \emptyc & \cellcolor{oksky!32}\halfc & \emptyc & \emptyc & \emptyc & \cellcolor{oksky!32}\halfc & \emptyc & \cellcolor{oksky!32}\halfc & \cellcolor{oksky!32}\halfc & \emptyc \\
19 & {\sffamily\fontsize{8.9}{9.5}\selectfont deribit agent (Vida)}~\cite{repo_deribitagent} & A2 & \cellcolor{okverm!16}W4 & \cellcolor{okverm!80}\textcolor{white}{\textbf{unb.}} & user & \cellcolor{okblue!62}\textcolor{white}{\fullc} & \cellcolor{oksky!32}\halfc & \cellcolor{oksky!32}\halfc & \cellcolor{oksky!32}\halfc & \cellcolor{oksky!32}\halfc & \emptyc & \cellcolor{oksky!32}\halfc & \cellcolor{oksky!32}\halfc & \emptyc & \emptyc & \cellcolor{oksky!32}\halfc & \emptyc & \emptyc & \emptyc & \cellcolor{oksky!32}\halfc & \cellcolor{okblue!62}\textcolor{white}{\fullc} & \cellcolor{oksky!32}\halfc & \cellcolor{oksky!32}\halfc & \emptyc \\
20 & {\sffamily\fontsize{8.9}{9.5}\selectfont options-trading-bot}~\cite{repo_rohan5commit} & \cellcolor{okverm!16}A4 & W3 & \cellcolor{okorange!45}asgn. & user & \cellcolor{oksky!32}\halfc & \cellcolor{oksky!32}\halfc & \cellcolor{oksky!32}\halfc & \cellcolor{oksky!32}\halfc & \cellcolor{oksky!32}\halfc & \cellcolor{oksky!32}\halfc & \cellcolor{oksky!32}\halfc & \cellcolor{oksky!32}\halfc & \emptyc & \emptyc & \emptyc & \emptyc & \cellcolor{oksky!32}\halfc & \emptyc & \emptyc & \cellcolor{okblue!62}\textcolor{white}{\fullc} & \cellcolor{okblue!62}\textcolor{white}{\fullc} & \cellcolor{oksky!32}\halfc & \emptyc \\
21 & {\sffamily\fontsize{8.9}{9.5}\selectfont Agentic-Trader}~\cite{repo_agentictrader} & \cellcolor{okverm!16}A3 & W3 & \cellcolor{okorange!45}asgn. & user & \cellcolor{oksky!32}\halfc & \cellcolor{okblue!62}\textcolor{white}{\fullc} & \cellcolor{okblue!62}\textcolor{white}{\fullc} & \cellcolor{okblue!62}\textcolor{white}{\fullc} & \cellcolor{oksky!32}\halfc & \cellcolor{oksky!32}\halfc & \cellcolor{oksky!32}\halfc & \cellcolor{okblue!62}\textcolor{white}{\fullc} & \cellcolor{oksky!32}\halfc & \cellcolor{oksky!32}\halfc & \cellcolor{okblue!62}\textcolor{white}{\fullc} & \cellcolor{oksky!32}\halfc & \cellcolor{oksky!32}\halfc & \cellcolor{okblue!62}\textcolor{white}{\fullc} & \cellcolor{oksky!32}\halfc & \cellcolor{okblue!62}\textcolor{white}{\fullc} & \cellcolor{oksky!32}\halfc & \cellcolor{okblue!62}\textcolor{white}{\fullc} & \cellcolor{oksky!32}\halfc \\
22 & {\sffamily\fontsize{8.9}{9.5}\selectfont agentic-trading-research}~\cite{repo_agentictradingresearch} & \cellcolor{okverm!16}A4 & W1 & \cellcolor{okyellow!55}prem. & user & \cellcolor{oksky!32}\halfc & \cellcolor{oksky!32}\halfc & \cellcolor{oksky!32}\halfc & \cellcolor{oksky!32}\halfc & \cellcolor{okblue!62}\textcolor{white}{\fullc} & \cellcolor{oksky!32}\halfc & \cellcolor{oksky!32}\halfc & \cellcolor{oksky!32}\halfc & \cellcolor{oksky!32}\halfc & \emptyc & \emptyc & \emptyc & \emptyc & \cellcolor{oksky!32}\halfc & \emptyc & \cellcolor{okblue!62}\textcolor{white}{\fullc} & \cellcolor{oksky!32}\halfc & \cellcolor{oksky!32}\halfc & \emptyc \\
23 & {\sffamily\fontsize{8.9}{9.5}\selectfont alpaca-options-agent}~\cite{repo_alpacaoptionsagent} & A1 & W2 & none & user & \cellcolor{okblue!62}\textcolor{white}{\fullc} & \cellcolor{okblue!62}\textcolor{white}{\fullc} & \cellcolor{okblue!62}\textcolor{white}{\fullc} & \cellcolor{okblue!62}\textcolor{white}{\fullc} & \cellcolor{okblue!62}\textcolor{white}{\fullc} & \cellcolor{oksky!32}\halfc & \cellcolor{oksky!32}\halfc & \cellcolor{okblue!62}\textcolor{white}{\fullc} & \emptyc & \emptyc & \cellcolor{oksky!32}\halfc & \emptyc & \emptyc & \emptyc & \cellcolor{oksky!32}\halfc & \cellcolor{okblue!62}\textcolor{white}{\fullc} & \cellcolor{okblue!62}\textcolor{white}{\fullc} & \cellcolor{okblue!62}\textcolor{white}{\fullc} & \emptyc \\
24 & {\sffamily\fontsize{8.9}{9.5}\selectfont Theta Desk}~\cite{repo_thetadesk} & A1 & \cellcolor{okverm!16}W4 & none & user & \cellcolor{okblue!62}\textcolor{white}{\fullc} & \cellcolor{oksky!32}\halfc & \cellcolor{oksky!32}\halfc & \cellcolor{oksky!32}\halfc & \cellcolor{okblue!62}\textcolor{white}{\fullc} & \cellcolor{oksky!32}\halfc & \cellcolor{oksky!32}\halfc & \cellcolor{okblue!62}\textcolor{white}{\fullc} & \emptyc & \emptyc & \emptyc & \emptyc & \cellcolor{oksky!32}\halfc & \emptyc & \cellcolor{oksky!32}\halfc & \cellcolor{okblue!62}\textcolor{white}{\fullc} & \cellcolor{oksky!32}\halfc & \cellcolor{okblue!62}\textcolor{white}{\fullc} & \emptyc \\
25 & {\sffamily\fontsize{8.9}{9.5}\selectfont AI-trading-Agent}~\cite{repo_liyang} & A0 & W0 & none & user & \cellcolor{okblue!62}\textcolor{white}{\fullc} & \emptyc & \emptyc & \emptyc & \cellcolor{oksky!32}\halfc & \emptyc & \cellcolor{oksky!32}\halfc & \cellcolor{oksky!32}\halfc & \cellcolor{oksky!32}\halfc & \emptyc & \emptyc & \emptyc & \cellcolor{oksky!32}\halfc & \cellcolor{oksky!32}\halfc & \emptyc & \cellcolor{black!9}\textcolor{black!55}{--} & \cellcolor{oksky!32}\halfc & \cellcolor{black!9}\textcolor{black!55}{--} & \emptyc \\
26 & {\sffamily\fontsize{8.9}{9.5}\selectfont news-agent}~\cite{repo_newsagent} & A0 & W0 & none & user & \cellcolor{okblue!62}\textcolor{white}{\fullc} & \cellcolor{oksky!32}\halfc & \cellcolor{oksky!32}\halfc & \emptyc & \cellcolor{oksky!32}\halfc & \emptyc & \cellcolor{oksky!32}\halfc & \emptyc & \cellcolor{oksky!32}\halfc & \emptyc & \emptyc & \emptyc & \emptyc & \emptyc & \cellcolor{oksky!32}\halfc & \cellcolor{black!9}\textcolor{black!55}{--} & \cellcolor{oksky!32}\halfc & \cellcolor{black!9}\textcolor{black!55}{--} & \emptyc \\
\midrule
\multicolumn{2}{@{}l}{\textbf{Count} (of \nSchemes{}; \fullc{} per column)} & \nAutoExec{}\,A3+ & \nWfour{}\,W4 & \nHunbounded{}\,unb. & \nHtwoPlatform{}\,plat. & \nHthreeInterceptable{}\,\fullc & \nRoneFull{} & \nRtwoFull{} & \nRthreeFull{} & \nRfourFull{} & \nRfiveFull{} & \nRsixFull{} & \nRsevenFull{} & \nConeFull{} & \nCtwoFull{} & \nCthreeFull{} & \nCfourFull{} & \nCfiveFull{} & \nCsixFull{} & \nCsevenFull{} & \nCeightFull{} & \nCnineFull{} & \nCtenFull{} & \nCelevenFull{} \\
\bottomrule
\end{tabular}
\end{table*}

Table~\ref{tab:schemes} gives every score, and counts are over all \nSchemes{} scored systems unless noted.

\subsection{RQ1: A Young Ecosystem Built Around One Event}
\label{sec:rq1}

\begin{figure}[!tb]
\centering
\includegraphics[width=\columnwidth]{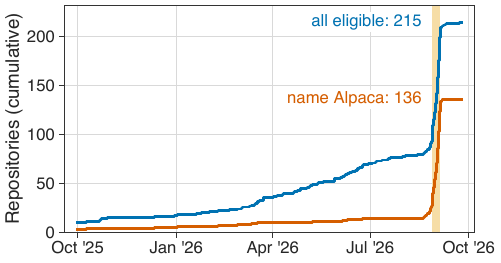}
\caption{\textbf{Eligible options-agent repositories by creation date.} Cumulative count of the \nEligible{} eligible GitHub repositories (blue) and of the \nEligibleAlpaca{} that name Alpaca in their own name, description, or topics (orange), from October 2025 to the snapshot; \timelineBefore{} repositories created before October 2025 start the blue line. The shaded band is the Alpaca AI Trading Agents Hackathon (late August to early September 2026)~\cite{lablab2026alpacahackathon}. Most of the population appeared within days of that event.}
\label{fig:timeline}
\end{figure}

The open-source population of LLM options agents is young, has little reach, and clusters around one event and one broker. Of the \nEligible{} eligible repositories, \nEligibleTwentySix{} were created in 2026 and \nEligibleBeforeTwentySix{} earlier. Figure~\ref{fig:timeline} shows steady growth through mid-2026 followed by a jump. In the two weeks around the end of August 2026, \nEligibleTwoWeek{} repositories (\pctEligibleTwoWeek{}) were created. Those two weeks bracket the Alpaca AI Trading Agents Hackathon, an online event in late August and early September 2026 whose rules require every strategy to trade options~\cite{lablab2026alpacahackathon}, and \nEligibleHackathon{} repositories name a hackathon in their own name, description, or topics. Thus, most of the population exists because an event asked for it, and much of it is likely to go unmaintained once the event ends.

The same event concentrated the population on one broker. Of the \nEligible{} repositories, \nEligibleAlpaca{} name Alpaca, \nEligibleTwoWeekAlpaca{} of them created inside the two-week window. Because Alpaca rejects uncovered options at every options level~\cite{alpaca2026optionsdocs}, the broker's policy bounds this population at assignment, whatever the agents themselves check. By contrast, agent code written against a write-enabled broker inherits that broker's scope. For example, schwab-mcp and tasty-agent accept sell-to-open legs with no coverage check, so their authority equals the options level of the account they connect to (Appendix~\ref{app:schemes}). The same code is bounded at assignment on one account and unbounded on another.

Assignment is a bound, not a small loss. When a cash-secured put is assigned, the account buys 100 shares per contract at the strike, so a fall in the underlying passes to the account in full, up to the strike times 100 less the premium received. A covered call that is assigned sells shares the account wanted to keep. The hackathon population therefore cannot lose without limit through Alpaca, yet a manipulated put can still turn its cash into a falling stock position (O-5).

\rih{Designs from the event.} The event produced careful designs as well as careless ones. One hackathon submission in our scored sample, alpaca-options-agent, runs a deterministic risk gate before the model and lets the model only decline a candidate the gate approved (A1). It submits each spread as one native multi-leg order, so its structure cannot leave an uncovered leg, and it enforces eight safeguards outside the model, as many as any system we scored, on a paper account with no stars. Its inputs, however, are as unchecked as everyone else's, with nothing documented for source authentication (C1), market-data integrity (C2) or instruction--data separation (C4) (Appendix~\ref{app:schemes}). The event thus produced at least one design that bounds the order, and none that verifies the text behind it.

Reach is concentrated in a few projects. \nEligibleZeroStar{} of the \nEligible{} repositories (\pctEligibleZeroStar{}) have no stars, and \nEligibleTwentyStars{} in the union have at least \starThreshold{}. Although the most-starred general LLM trading frameworks we checked have up to 108,459 stars, none of them supports options (Appendix~\ref{app:search}). Adoption therefore rests on general trading platforms that expose options to agents, such as openalgo (\starsOpenalgo{} stars) and lumibot (\starsLumibot{} stars), and on the \nBroker{} broker interfaces, each of which reaches every customer who enables it.

\begin{figure}[!tb]
\centering
\includegraphics[width=\columnwidth]{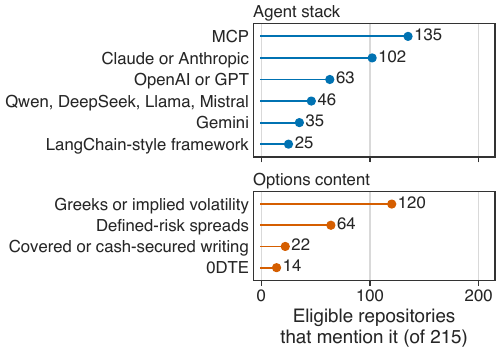}
\caption{\textbf{Term mentions in the eligible repositories.} For each term group, the number of the \nEligible{} eligible repositories whose name, description, topics, or README mention it, from the keyword matches logged during screening. A repository can mention several groups. The open-weight row counts Qwen, DeepSeek, Llama, and Mistral, and the framework row counts LangChain, LangGraph, CrewAI, and AutoGen.}
\label{fig:mentions}
\end{figure}

The population builds on the broker interfaces we score and on a few model vendors (Figure~\ref{fig:mentions}). Of the \nEligible{} repositories, \nMentionMCP{} mention MCP in their name, description, topics, or README, so their agents read chain data and send orders through the tool channel that tool poisoning targets~\cite{mcptox2026}. \nMentionClaude{} mention Claude or Anthropic, \nMentionOpenAI{} OpenAI or GPT, \nMentionOpenWeight{} an open-weight family, and \nMentionGemini{} Gemini, while \nMentionMultiVendor{} mention models from at least two of these vendors. On the options side, \nMentionGreeksIV{} mention Greeks or implied volatility, the derived numbers that no scored system cross-checks against an independent feed (\S\ref{sec:rq3}). Another \nMentionSpreads{} mention defined-risk spreads, the structures whose bound O-1 removes, and \nMentionWriting{} mention covered or cash-secured writing, the short positions that keep O-5 reachable on brokers that ban uncovered writing.

\begin{takeaway}
\textbf{RQ1.} LLM options agents are a 2026 phenomenon. \nEligibleTwentySix{} of the \nEligible{} open-source repositories were created this year, \pctEligibleTwoWeek{} of them around one hackathon, and \nEligibleAlpaca{} name a single broker, whose write policy sets their worst case.
\end{takeaway}

\subsection{RQ2: The Worst Case Is Reachable}
\label{sec:rq2}

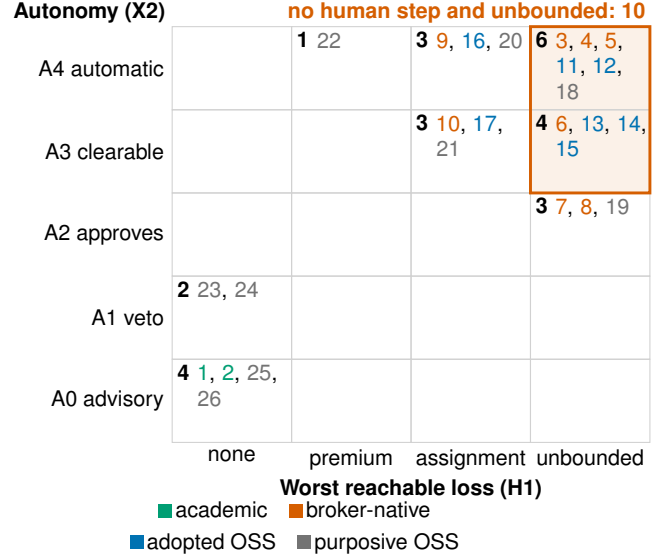
\begin{figure}[!tb]
\centering
\begin{tikzpicture}[font=\sffamily\small]
\fill[okverm!9] (4.74,3.30) rectangle (6.32,5.50);
\draw[black!18, line width=0.4pt] (0.00,0.00) rectangle (1.58,1.10);
\node[anchor=north west, inner sep=1.5pt, font=\sffamily\small\bfseries] at (0.01,1.09) {4};
\node[anchor=north west, inner sep=0pt, text width=1.16cm, align=left, font=\sffamily\fontsize{9}{9.6}\selectfont] at (0.33,1.03) {\textcolor{okgreen}{1}, \textcolor{okgreen}{2}, \textcolor{black!55}{25}, \textcolor{black!55}{26}};
\draw[black!18, line width=0.4pt] (0.00,1.10) rectangle (1.58,2.20);
\node[anchor=north west, inner sep=1.5pt, font=\sffamily\small\bfseries] at (0.01,2.19) {2};
\node[anchor=north west, inner sep=0pt, text width=1.16cm, align=left, font=\sffamily\fontsize{9}{9.6}\selectfont] at (0.33,2.13) {\textcolor{black!55}{23}, \textcolor{black!55}{24}};
\draw[black!18, line width=0.4pt] (0.00,2.20) rectangle (1.58,3.30);
\draw[black!18, line width=0.4pt] (0.00,3.30) rectangle (1.58,4.40);
\draw[black!18, line width=0.4pt] (0.00,4.40) rectangle (1.58,5.50);
\draw[black!18, line width=0.4pt] (1.58,0.00) rectangle (3.16,1.10);
\draw[black!18, line width=0.4pt] (1.58,1.10) rectangle (3.16,2.20);
\draw[black!18, line width=0.4pt] (1.58,2.20) rectangle (3.16,3.30);
\draw[black!18, line width=0.4pt] (1.58,3.30) rectangle (3.16,4.40);
\draw[black!18, line width=0.4pt] (1.58,4.40) rectangle (3.16,5.50);
\node[anchor=north west, inner sep=1.5pt, font=\sffamily\small\bfseries] at (1.59,5.49) {1};
\node[anchor=north west, inner sep=0pt, text width=1.16cm, align=left, font=\sffamily\fontsize{9}{9.6}\selectfont] at (1.91,5.43) {\textcolor{black!55}{22}};
\draw[black!18, line width=0.4pt] (3.16,0.00) rectangle (4.74,1.10);
\draw[black!18, line width=0.4pt] (3.16,1.10) rectangle (4.74,2.20);
\draw[black!18, line width=0.4pt] (3.16,2.20) rectangle (4.74,3.30);
\draw[black!18, line width=0.4pt] (3.16,3.30) rectangle (4.74,4.40);
\node[anchor=north west, inner sep=1.5pt, font=\sffamily\small\bfseries] at (3.17,4.39) {3};
\node[anchor=north west, inner sep=0pt, text width=1.16cm, align=left, font=\sffamily\fontsize{9}{9.6}\selectfont] at (3.49,4.33) {\textcolor{okverm}{10}, \textcolor{okblue}{17}, \textcolor{black!55}{21}};
\draw[black!18, line width=0.4pt] (3.16,4.40) rectangle (4.74,5.50);
\node[anchor=north west, inner sep=1.5pt, font=\sffamily\small\bfseries] at (3.17,5.49) {3};
\node[anchor=north west, inner sep=0pt, text width=1.16cm, align=left, font=\sffamily\fontsize{9}{9.6}\selectfont] at (3.49,5.43) {\textcolor{okverm}{9}, \textcolor{okblue}{16}, \textcolor{black!55}{20}};
\draw[black!18, line width=0.4pt] (4.74,0.00) rectangle (6.32,1.10);
\draw[black!18, line width=0.4pt] (4.74,1.10) rectangle (6.32,2.20);
\draw[black!18, line width=0.4pt] (4.74,2.20) rectangle (6.32,3.30);
\node[anchor=north west, inner sep=1.5pt, font=\sffamily\small\bfseries] at (4.75,3.29) {3};
\node[anchor=north west, inner sep=0pt, text width=1.16cm, align=left, font=\sffamily\fontsize{9}{9.6}\selectfont] at (5.07,3.23) {\textcolor{okverm}{7}, \textcolor{okverm}{8}, \textcolor{black!55}{19}};
\draw[black!18, line width=0.4pt] (4.74,3.30) rectangle (6.32,4.40);
\node[anchor=north west, inner sep=1.5pt, font=\sffamily\small\bfseries] at (4.75,4.39) {4};
\node[anchor=north west, inner sep=0pt, text width=1.16cm, align=left, font=\sffamily\fontsize{9}{9.6}\selectfont] at (5.07,4.33) {\textcolor{okverm}{6}, \textcolor{okblue}{13}, \textcolor{okblue}{14}, \textcolor{okblue}{15}};
\draw[black!18, line width=0.4pt] (4.74,4.40) rectangle (6.32,5.50);
\node[anchor=north west, inner sep=1.5pt, font=\sffamily\small\bfseries] at (4.75,5.49) {6};
\node[anchor=north west, inner sep=0pt, text width=1.16cm, align=left, font=\sffamily\fontsize{9}{9.6}\selectfont] at (5.07,5.43) {\textcolor{okverm}{3}, \textcolor{okverm}{4}, \textcolor{okverm}{5}, \textcolor{okblue}{11}, \textcolor{okblue}{12}, \textcolor{black!55}{18}};
\draw[okverm, line width=1.1pt] (4.74,3.30) rectangle (6.32,5.50);
\node[anchor=south east, inner sep=1.5pt, text=okverm, font=\sffamily\small\bfseries] at (6.32,5.50) {no human step and unbounded: \nUnboundedAuto{}};
\node[anchor=north, inner sep=1.5pt] at (0.79,-0.03) {none};
\node[anchor=north, inner sep=1.5pt] at (2.37,-0.03) {premium};
\node[anchor=north, inner sep=1.5pt] at (3.95,-0.03) {assignment};
\node[anchor=north, inner sep=1.5pt] at (5.53,-0.03) {unbounded};
\node[anchor=east, inner sep=2pt] at (-0.03,0.55) {A0 advisory};
\node[anchor=east, inner sep=2pt] at (-0.03,1.65) {A1 veto};
\node[anchor=east, inner sep=2pt] at (-0.03,2.75) {A2 approves};
\node[anchor=east, inner sep=2pt] at (-0.03,3.85) {A3 clearable};
\node[anchor=east, inner sep=2pt] at (-0.03,4.95) {A4 automatic};
\node[anchor=north, font=\sffamily\small\bfseries] at (3.16,-0.36) {Worst reachable loss (H1)};
\node[anchor=south east, inner sep=1.5pt, font=\sffamily\small\bfseries] at (-0.03,5.50) {Autonomy (X2)};
\node[anchor=north, inner sep=0pt, align=center] at (1.62,-0.78) {\textcolor{okgreen}{\rule{5pt}{5pt}}\,academic\quad \textcolor{okverm}{\rule{5pt}{5pt}}\,broker-native\\[1pt] \textcolor{okblue}{\rule{5pt}{5pt}}\,adopted OSS\quad \textcolor{black!55}{\rule{5pt}{5pt}}\,purposive OSS};
\end{tikzpicture}
\caption{\textbf{Positions of the \nSchemes{} systems.} Each cell pairs the worst loss the documented order path can reach (H1) with an autonomy level (X2) and gives the number of systems and their row numbers in Table~\ref{tab:schemes}, colored by sampling stratum. The outlined block holds the \nUnboundedAuto{} systems that can reach unbounded loss behind an agent-clearable gate or none.}
\label{fig:landscape}
\end{figure}

\begin{table}[t]
\centering
\footnotesize
\caption{Exposure and safeguards by sampling stratum. Each cell counts the systems in the stratum that have an order path, can execute with no mandatory human step (A3 or A4), can write uncovered shorts (W4), can reach unbounded loss (H1), enforce a pre-trade limit outside the model (C8 full), and authenticate sources or cross-check market data (C1 or C2 full). The purposive stratum describes designs, and prevalence is read from the other rows.}
\label{tab:strata}
\setlength{\tabcolsep}{2.6pt}
\begin{tabular}{@{}l rrrrrrr@{}}
\toprule
Stratum & \makecell[r]{Sys-\\tems} & \makecell[r]{Can\\order} & \makecell[r]{No\\human} & \makecell[r]{Un-\\cov.} & \makecell[r]{Unb.\\loss} & \makecell[r]{Size\\limit} & \makecell[r]{Input\\check} \\
\midrule
Academic & 2 & 0 & 0 & 0 & 0 & 0 & 0 \\
Broker-native & 8 & 8 & 6 & 6 & 6 & 3 & 0 \\
Open source, $\geq$\starThreshold{} stars & 7 & 7 & 7 & 5 & 5 & 1 & 0 \\
Open source, purposive & 9 & 7 & 4 & 3 & 2 & 6 & 0 \\
\midrule
\textbf{All systems} & \textbf{26} & \textbf{22} & \textbf{17} & \textbf{14} & \textbf{13} & \textbf{10} & \textbf{0} \\
\bottomrule
\end{tabular}
\end{table}

Most systems that can trade do so with no human in the loop, and half of the scored systems can reach unbounded loss.

\rih{Autonomy.} Of the \nSchemes{} systems, \nOrderCapable{} have an order path and \nAutoExec{} can execute without a mandatory human step on each order, \nAfour{} with no gate (A4) and \nAthree{} behind a gate that the agent can clear, that is off by default, or that a documented setting removes (A3). For example, tastytrade lets the agent that requests a dry-run token also redeem it~\cite{tastytrademcp2026}, and Robinhood's approval step is skipped when the customer asks for it~\cite{robinhoodsupport2026}. Just \nAtwo{} systems require a human to approve every order in every documented configuration.

\rih{Write scope and worst loss.} \nWfour{} systems can place an uncovered short through their documented order path (W4), and \nHunbounded{} reach unbounded loss (H1). Because its LLM can only veto (A1), Theta Desk reaches no loss by itself despite its W4 scope. Another \nHassignment{} systems are bounded by assignment, among them Robinhood and Alpaca, whose brokers offer covered calls, cash-secured puts, and defined-risk spreads with no uncovered writing~\cite{robinhoodmcp2026,robinhoodlevels2026,alpaca2026optionsdocs}. Figure~\ref{fig:landscape} places each system by autonomy and reachable loss. Its outlined block holds the \nUnboundedAuto{} systems in which an agent-clearable gate, or no gate at all, meets unbounded loss.

\rih{Strata.} Table~\ref{tab:strata} breaks the profile down by sampling stratum, and broker interfaces are the most exposed stratum. Of the \nBroker{}, \nBrokerUnbounded{} reach unbounded loss and \nBrokerAutoExec{} execute without a mandatory human step, compared with \nOSSUnbounded{} and \nOSSAutoExec{} of the \nOSS{} open-source systems. \nBrokerUnboundedNoLimit{} broker interfaces, OKX Agent Trade Kit, Public MCP, and Webull MCP, combine no gate, uncovered writing, and no pre-trade limit outside the model. Similarly, the census strata give the same picture, with \nCensusUnbounded{} of the \nCensus{} census systems reaching unbounded loss.

\rih{Unbounded order paths.} These three broker interfaces fail in three plain ways (Appendix~\ref{app:schemes}). OKX Agent Trade Kit's order tool posts straight to the exchange with no confirmation, dry-run or preview parameter, and writing an option there is an ordinary margined sell. Public MCP builds each order from the raw tool arguments and accepts sell-to-open legs after a syntax check. Webull MCP ships notional and quantity risk controls in its code, but neither of its option-order validators calls them. In each case one tool call can open a naked short whose loss has no ceiling, on whatever account the customer connected to the agent. Open-source servers repeat the pattern at the account level. tasty-agent and schwab-mcp accept sell-to-open legs with no coverage check, so the agent's authority equals the options level of the account it is given, and schwab-mcp's human approval step disappears behind a documented command-line flag.

\rih{Human approval.} Two broker products, IBKR's AI integration and the TradeStation connector, require the customer to approve every order (A2), and both still reach unbounded loss. The orders they draft can be any order the account is approved for, IBKR's Level~4 permits naked calls, and TradeStation's connector states that it never narrows the account's permissions. A customer who approves a credit spread that has lost its protective leg sees a routine ticket that collects a premium, which is exactly the order O-1 induces (\S\ref{sec:primitives}). Of the \nSchemes{} systems, \nAtwo{} require such an approval in every configuration, so human review is rare, and where it exists it bounds how often the agent acts rather than how much one order can lose.

\rih{Blast radius.} \nHtwoPlatform{} of the \nSchemes{} systems have a platform-scale blast radius (H2), so one design's failure reaches every user of the platform. They are all \nBroker{} broker interfaces and the \nOSSAdopted{} adopted open-source projects, and \nPlatformUnbounded{} of them can reach unbounded loss.

\begin{takeaway}
\textbf{RQ2.} \nAutoExec{} of the \nSchemes{} systems can execute with no mandatory human step, and \nHunbounded{} can reach unbounded loss. Broker interfaces are the most exposed. \nBrokerUnbounded{} of \nBroker{} reach unbounded loss, and \nBrokerUnboundedNoLimit{} of them combine no gate, uncovered writing, and no out-of-model limit.
\end{takeaway}

\subsection{RQ3: Limits on Size, No Checks on Inputs}
\label{sec:rq3}

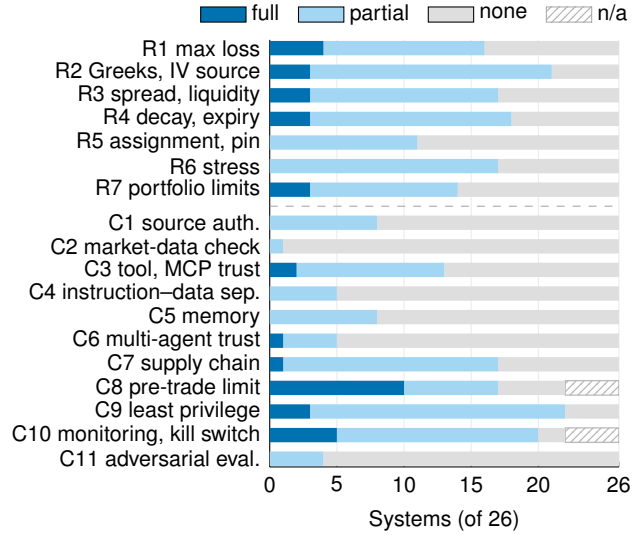
\begin{figure}[!tb]
\centering
\begin{tikzpicture}
\begin{axis}[
  xbar stacked, bar width=5.4pt, y dir=reverse, scale only axis,
  width=4.62cm, height=5.65cm, xmin=0, xmax=26,
  xtick={0,5,10,15,20,26}, ytick={1.0,2.0,3.0,4.0,5.0,6.0,7.0,8.4,9.4,10.4,11.4,12.4,13.4,14.4,15.4,16.4,17.4,18.4}, yticklabels={{R1 max loss},{R2 Greeks, IV source},{R3 spread, liquidity},{R4 decay, expiry},{R5 assignment, pin},{R6 stress},{R7 portfolio limits},{C1 source auth.},{C2 market-data check},{C3 tool, MCP trust},{C4 instruction--data sep.},{C5 memory},{C6 multi-agent trust},{C7 supply chain},{C8 pre-trade limit},{C9 least privilege},{C10 monitoring, kill switch},{C11 adversarial eval.}},
  tick label style={font=\sffamily\small}, label style={font=\sffamily\small},
  y tick label style={anchor=east}, major tick length=2pt, ytick style={draw=none},
  xticklabel={\sffamily\pgfmathprintnumber[assume math mode=true]{\tick}},
  xlabel={Systems (of \nSchemes{})}, axis x line*=bottom, axis y line*=left,
  enlarge y limits={abs=0.35}, xmajorgrids, grid style={draw=black!8},
  legend style={font=\sffamily\small, at={(0.42,1.01)}, anchor=south, legend columns=4,
    draw=none, /tikz/every even column/.append style={column sep=5pt}}, area legend,
]
\addplot[fill=okblue, draw=none] coordinates {(4,1.0) (3,2.0) (3,3.0) (3,4.0) (0,5.0) (0,6.0) (3,7.0) (0,8.4) (0,9.4) (2,10.4) (0,11.4) (0,12.4) (1,13.4) (1,14.4) (10,15.4) (3,16.4) (5,17.4) (0,18.4)};
\addplot[fill=oksky!55, draw=none] coordinates {(12,1.0) (18,2.0) (14,3.0) (15,4.0) (11,5.0) (17,6.0) (11,7.0) (8,8.4) (1,9.4) (11,10.4) (5,11.4) (8,12.4) (4,13.4) (16,14.4) (7,15.4) (19,16.4) (15,17.4) (4,18.4)};
\addplot[fill=black!13, draw=none] coordinates {(10,1.0) (5,2.0) (9,3.0) (8,4.0) (15,5.0) (9,6.0) (12,7.0) (18,8.4) (25,9.4) (13,10.4) (21,11.4) (18,12.4) (21,13.4) (9,14.4) (5,15.4) (4,16.4) (2,17.4) (22,18.4)};
\addplot[fill=white, draw=black!35, line width=0.3pt, pattern=north east lines, pattern color=black!30] coordinates {(0,1.0) (0,2.0) (0,3.0) (0,4.0) (0,5.0) (0,6.0) (0,7.0) (0,8.4) (0,9.4) (0,10.4) (0,11.4) (0,12.4) (0,13.4) (0,14.4) (4,15.4) (0,16.4) (4,17.4) (0,18.4)};
\legend{full, partial, none, n/a}
\draw[black!35, dashed, line width=0.4pt] (axis cs:0,7.7) -- (axis cs:26,7.7);
\end{axis}
\end{tikzpicture}
\caption{\textbf{Safeguard coverage.} For each robustness (R1--R7) and security (C1--C11) indicator, the number of the \nSchemes{} systems that score it full (enforced outside the model), partial (model-dependent, optional, or clearable by the agent), or none documented, with hatching where the indicator does not apply because the system has no order path (C8, C10). R5, R6, C1, C2, C4, C5, and C11 are full in none of the systems, and the most common full safeguard is C8 (pre-trade limit, \nCeightFull{} of the \nOrderCapable{} systems that can place orders).}
\label{fig:coverage}
\end{figure}

\begin{figure}[!tb]
\centering
\begin{tikzpicture}[font=\sffamily\small]
\draw[okverm, line width=0.8pt] (2.49,0.72) -- (4.27,0.72);
\node[anchor=south, inner sep=1pt, text=okverm, font=\sffamily\small\bfseries] at (3.38,0.74) {stop the text};
\draw[okblue, line width=0.8pt] (4.35,0.72) -- (7.37,0.72);
\node[anchor=south, inner sep=1pt, text=okblue, font=\sffamily\small\bfseries] at (5.86,0.74) {catch the error};
\draw[black!60, line width=0.8pt] (7.45,0.72) -- (7.99,0.72);
\node[anchor=south, inner sep=1pt, text=black!60, font=\sffamily\small\bfseries] at (7.72,0.74) {cap size};
\node[anchor=south, inner sep=1pt, font=\sffamily\small\bfseries] at (2.76,0.18) {C1};
\node[anchor=south, inner sep=1pt, font=\sffamily\small\bfseries] at (3.38,0.18) {C2};
\node[anchor=south, inner sep=1pt, font=\sffamily\small\bfseries] at (4.00,0.18) {C4};
\node[anchor=south, inner sep=1pt, font=\sffamily\small\bfseries] at (4.62,0.18) {R1};
\node[anchor=south, inner sep=1pt, font=\sffamily\small\bfseries] at (5.24,0.18) {R2};
\node[anchor=south, inner sep=1pt, font=\sffamily\small\bfseries] at (5.86,0.18) {R3};
\node[anchor=south, inner sep=1pt, font=\sffamily\small\bfseries] at (6.48,0.18) {R4};
\node[anchor=south, inner sep=1pt, font=\sffamily\small\bfseries] at (7.10,0.18) {R5};
\node[anchor=south, inner sep=1pt, font=\sffamily\small\bfseries] at (7.72,0.18) {C8};
\fill[black!4] (0,0.21) rectangle (8.03,-0.21);
\node[anchor=west, inner sep=0pt] at (0.02,0.00) {\textbf{O-1}\ risk structure};
\fill[black!70] (2.76,0.00) circle (0.085);
\fill[black!70] (3.38,0.00) circle (0.085);
\fill[black!70] (4.00,0.00) circle (0.085);
\fill[black!70] (4.62,0.00) circle (0.085);
\fill[black!70] (7.10,0.00) circle (0.085);
\fill[black!70] (7.72,0.00) circle (0.085);
\node[anchor=west, inner sep=0pt] at (0.02,-0.43) {\textbf{O-2}\ strike, expiry};
\fill[black!70] (2.76,-0.43) circle (0.085);
\fill[black!70] (5.86,-0.43) circle (0.085);
\fill[black!70] (6.48,-0.43) circle (0.085);
\fill[black!70] (7.72,-0.43) circle (0.085);
\fill[black!4] (0,-0.65) rectangle (8.03,-1.07);
\node[anchor=west, inner sep=0pt] at (0.02,-0.86) {\textbf{O-3}\ Greeks, IV};
\fill[black!70] (2.76,-0.86) circle (0.085);
\fill[black!70] (3.38,-0.86) circle (0.085);
\fill[black!70] (5.24,-0.86) circle (0.085);
\fill[black!70] (6.48,-0.86) circle (0.085);
\fill[black!70] (7.72,-0.86) circle (0.085);
\node[anchor=west, inner sep=0pt] at (0.02,-1.29) {\textbf{O-4}\ expiry pressure};
\fill[black!70] (2.76,-1.29) circle (0.085);
\fill[black!70] (6.48,-1.29) circle (0.085);
\fill[black!70] (7.72,-1.29) circle (0.085);
\fill[black!4] (0,-1.50) rectangle (8.03,-1.94);
\node[anchor=west, inner sep=0pt] at (0.02,-1.72) {\textbf{O-5}\ assignment, pin};
\fill[black!70] (2.76,-1.72) circle (0.085);
\fill[black!70] (4.00,-1.72) circle (0.085);
\fill[black!70] (4.62,-1.72) circle (0.085);
\fill[black!70] (7.10,-1.72) circle (0.085);
\fill[black!70] (7.72,-1.72) circle (0.085);
\node[anchor=west, inner sep=0pt] at (0.02,-2.15) {\textbf{O-6}\ profile, harvest};
\fill[black!70] (2.76,-2.15) circle (0.085);
\fill[black!70] (5.86,-2.15) circle (0.085);
\fill[black!70] (7.72,-2.15) circle (0.085);
\draw[black!40, line width=0.5pt] (0,-2.53) -- (8.03,-2.53);
\node[anchor=west, inner sep=0pt, text width=2.4cm, align=left] at (0.02,-2.78) {systems with it full};
\node[fill=okverm!75, text=white, minimum width=0.54cm, minimum height=0.36cm, inner sep=0pt, font=\sffamily\small\bfseries] at (2.76,-2.78) {\nConeFull{}};
\node[fill=okverm!75, text=white, minimum width=0.54cm, minimum height=0.36cm, inner sep=0pt, font=\sffamily\small\bfseries] at (3.38,-2.78) {\nCtwoFull{}};
\node[fill=okverm!75, text=white, minimum width=0.54cm, minimum height=0.36cm, inner sep=0pt, font=\sffamily\small\bfseries] at (4.00,-2.78) {\nCfourFull{}};
\node[fill=okblue!35, text=black, minimum width=0.54cm, minimum height=0.36cm, inner sep=0pt, font=\sffamily\small\bfseries] at (4.62,-2.78) {\nRoneFull{}};
\node[fill=okblue!35, text=black, minimum width=0.54cm, minimum height=0.36cm, inner sep=0pt, font=\sffamily\small\bfseries] at (5.24,-2.78) {\nRtwoFull{}};
\node[fill=okblue!35, text=black, minimum width=0.54cm, minimum height=0.36cm, inner sep=0pt, font=\sffamily\small\bfseries] at (5.86,-2.78) {\nRthreeFull{}};
\node[fill=okblue!35, text=black, minimum width=0.54cm, minimum height=0.36cm, inner sep=0pt, font=\sffamily\small\bfseries] at (6.48,-2.78) {\nRfourFull{}};
\node[fill=okverm!75, text=white, minimum width=0.54cm, minimum height=0.36cm, inner sep=0pt, font=\sffamily\small\bfseries] at (7.10,-2.78) {\nRfiveFull{}};
\node[fill=okblue!65, text=white, minimum width=0.54cm, minimum height=0.36cm, inner sep=0pt, font=\sffamily\small\bfseries] at (7.72,-2.78) {\nCeightFull{}};
\end{tikzpicture}
\caption{\textbf{Safeguards that would stop each primitive, with system counts.} A dot marks a safeguard that addresses a primitive. The robustness indicators follow the primitive's axes (R1--R5 operationalize A1--A5, and R3 also A6), the input-side controls follow its precursor classes (C1 for text sources, C2 for falsified tool fields, C4 for injected instructions), and the out-of-model limit C8 bounds every order. The bottom row counts the systems that score each safeguard full, out of \nSchemes{} (C8 out of the \nOrderCapable{} that can place orders). Only the size limit C8 is common, and a manipulated order within budget passes it.}
\label{fig:matrix}
\end{figure}

The safeguards we found decide whether an order is affordable and permitted. However, none of them decides whether the information behind the order was honest.

\rih{Input side.} Figure~\ref{fig:coverage} shows the input side as the least covered. No system scores full on source authentication (C1), market-data integrity (C2), instruction--data separation (C4), or adversarial evaluation (C11), and \nBrokerCtwoNone{} of the \nBroker{} broker interfaces document no market-data check at all. Of the \nOrderCapable{} order-capable systems, \nOrderNoFullC{} enforce none of the seven input-side controls (C1--C7) outside the model. These systems ingest the news, flow data, and Greeks that the primitives of \S\ref{sec:primitives} corrupt, and Figure~\ref{fig:matrix} shows that every primitive depends on at least one input-side control that no system enforces.

\rih{Order side.} A pre-trade limit enforced outside the model (C8) is the most common full safeguard, present in \nCeightFull{} of the \nOrderCapable{} order-capable systems, followed by monitoring with a kill switch (C10, \nCtenFull{}) and least privilege (C9, \nCnineFull{}). These limits are rarest where reach is largest (Table~\ref{tab:strata}). Among broker interfaces, \nBrokerCeightFull{} of \nBroker{} enforce C8, and among the adopted open-source projects \nOSSAdoptedCeightFull{} of \nOSSAdopted{} does, against \nPurposiveCeightFull{} of the \nPurposiveOrderCapable{} order-capable repositories in the purposive stratum. Since these controls bound size, a manipulated order that fits the budget passes all of them. A cap on order size also leaves the loss of an uncovered short unbounded, and \nCeightFullUnbounded{} of the \nHunbounded{} systems that reach unbounded loss enforce such a cap. Human review has the same blind spot, because a person approving a ticket sees the order and never sees the hidden instruction~\cite{greshake2023} or poisoned tool output~\cite{mcptox2026} behind it.

\rih{Options robustness.} Options-specific robustness is mostly partial. Assignment and pin handling (R5) and stress handling (R6) are full in no system, maximum-loss computation (R1) is full in \nRoneFull{}, and Greeks and IV provenance (R2) in \nRtwoFull{}. A partial score usually marks an instruction to the model, which catches an honest error and gives no protection against an attacker who writes into the same context.

\rih{Partial safeguards in practice.} Two examples show the pattern. tasty-agent's only confirmation requirement lives in a skill prompt, while its order tool executes on a single call with dry-run off by default. lumibot scores partial on seventeen of the eighteen safeguards, almost all of them rules the model is asked to follow or settings the operator may turn on. An attacker who writes into the agent's context writes into the same channel as these rules, so a partial score bounds an honest model's mistakes and leaves the induced order open.

\rih{The best-guarded systems.} The best-guarded systems are among the least adopted. Two repositories with zero and three stars, alpaca-options-agent and Agentic-Trader, each enforce eight safeguards outside the model, including a maximum-loss computation (R1) and a pre-trade limit (C8). By contrast, no broker interface enforces more than three, and the two most-starred open-source platforms, openalgo and lumibot, enforce two and none. Even the two best-guarded systems leave the inputs to the model, since neither enforces source authentication, market-data integrity or instruction--data separation outside it. Adoption has therefore gone to the designs that guard least, and none of the designs guards the text.

\begin{takeaway}
\textbf{RQ3.} \nInputVerified{} of the \nSchemes{} systems authenticate their sources or cross-check market data, and \nOrderNoFullC{} of the \nOrderCapable{} order-capable systems enforce no input-side control. The common safeguard is a size limit (\nCeightFull{} of \nOrderCapable{}), which a manipulated order that fits the budget passes.
\end{takeaway}

\section{Recommendations}
\label{sec:recommend}

Our findings point to the order as the place to bound an options agent's loss, because that is where the risk structure becomes concrete. Beyond FARSIGHT's case for invariants outside the model~\cite{wang2026soktradingagentsmarket}, options add a sharper reason. Because the manipulation we care about most asserts nothing false, the text gives a filter nothing to catch. Content-level detection looks for a malicious property of the input, such as an injected instruction, a fabricated figure, or a spoofed source, and a manipulation built from true statements has none. The harm arises when accurate text is mapped to a risk structure, since the same true earnings headline can justify a debit spread with a known maximum loss or an uncovered short call with none. The order is a small formal object whose worst-case loss at expiry is computable exactly~\cite{hull2021}, so defenses belong at the order boundary, with content filtering as a secondary layer. Brokers own that boundary, so the first five controls are theirs.

\rih{Defined-risk structures for agent accounts.} Agent-originated orders should be limited to structures with a finite maximum loss, with uncovered writing gated behind a separate approval that no agent tool call can grant. FINRA Rule 2360 requires each customer to be approved for specific option activities, including uncovered writing~\cite{finra2360}, and FINRA's 2021 action against Robinhood found the firm approving options accounts through ``option account approval bots'' under limited principal oversight~\cite{finra2021robinhoodawc}. Brokers should extend that approval to the agent as a distinct principal with its own options level. Robinhood's two official sources describe different agent scopes (\S\ref{sec:background}), and whichever one the broker enforces should be stated as the agent's level.

\rih{Per-order and per-leg maximum loss.} Before routing, the broker should compute the worst-case loss of the whole order and of each leg alone, then reject any order whose loss is unbounded or exceeds the account's cap. Today \nRoneFullOrder{} of the \nOrderCapable{} order-capable systems compute and enforce a maximum loss before entry (R1). The check matters per leg because a spread that is defined-risk as a package becomes naked if only its short leg fills, or if its long leg closes first.

\rih{Confirmation outside the agent.} The tastytrade MCP server binds submission to a dry-run check that an unattended agent can satisfy itself~\cite{tastytrademcp2026}, so the gate stops a careless agent and lets a steered one through. \nAthree{} of the systems we scored sit behind such an agent-clearable gate (\S\ref{sec:rq2}). A confirmation should travel through a channel the agent does not control, bind to a digest of the order, and show the broker-computed maximum loss.

\rih{Dedicated agent accounts.} Robinhood already places agents in a dedicated account with spending limits~\cite{robinhoodsupport2026}. The cap should also cover margin, and an agent account should never draw buying power from the owner's main account.

\rih{Pre-trade risk controls.} Under SEC Rule 15c3-5, a broker-dealer with market access must apply pre-trade credit, capital, and erroneous-order controls~\cite{sec2010rule15c3_5}. An agent MCP endpoint deserves the same filters: price collars against the quoted spread (A3), rejection of market orders on wide-spread contracts, limits on order size and frequency, and tighter limits near expiry, where decay and gamma dominate pricing (A4).

Agent builders control everything before the order reaches the broker, from the text the agent reads to the order it sends.

\rih{Untrusted text and tools.} Every news item, social post, and tool output should be handled as attacker-controllable, because indirect prompt injection needs no model access~\cite{greshake2023} and MCP tool descriptions are a documented poisoning vector~\cite{mcptox2026}. Signed feeds, C2PA manifests~\cite{c2pa2025}, and RFC 3161 timestamps~\cite{rfc3161} narrow who can write into the agent's context, although what is written still goes unjudged. Today no system we scored enforces source authentication (C1), and \nConePartial{} rely on an instruction to the model.

\rih{Numbers re-derived from the chain.} Greeks, implied volatility, and theoretical prices quoted in text should never enter a decision directly. The agent should recompute implied volatility from the quoted bid and ask by inverting a pricing model~\cite{blackscholes1973,hull2021}, derive Greeks from it, and discard any text-stated figure outside a tolerance, because neither the model nor a reviewer can check these numbers by eye (A2). Just \nRtwoFull{} of the \nSchemes{} systems derive Greeks and IV this way today (R2).

\rih{Separate reasoning and order channels.} CaMeL separates control flow from retrieved data~\cite{camel2025}, spotlighting marks input provenance~\cite{spotlighting2024}, and StruQ and SecAlign train models to ignore data-channel instructions~\cite{struq2025,secalign2025}. None of the systems we scored enforces such a separation outside the model (C4), and \nCfourPartial{} rely on prompt instructions. Even where deployed, these defenses stop injected instructions and leave accurate text that shifts judgment untouched, so the order-level checks remain necessary.

\rih{Sizing outside the model.} Position size, maximum loss per trade, and per-day loss limits belong in deterministic code between the model and the broker API, where a policy engine that the model cannot rewrite gates each order. \nCeightFull{} of the \nOrderCapable{} order-capable systems enforce such a limit today (C8), and in \nCeightPartial{} more the agent can clear or disable it.

\rih{Order-flow hygiene and liquidity limits.} Because an agent's orders are observable (A6), builders should avoid resting predictable orders in thin contracts, vary timing and size, and refuse contracts below an open-interest or volume floor. These controls lower what a profile-and-harvest attacker (O-6) can collect without having to detect the manipulation.

\rih{Adversarial testing before release.} No system documents a full adversarial evaluation (C11), and \nCelevenPartial{} document a partial one. Before going live, builders should replay the primitives of \S\ref{sec:primitives} through the agent's own order path on a paper-trading account and publish the worst loss of each resulting order with the code.

\section{Open Problems}
\label{sec:open}

Our measurement shows which safeguards options agents document and what their order paths can reach. Six problems remain open before a broker or a builder can show that an agent withstands the attacker of \S\ref{sec:threat}.

\begin{enumerate}[leftmargin=*,itemsep=2pt,topsep=2pt]
  \item \textbf{Attack success measured in dollars.} Because we scored designs from their public artifacts, the next measurement replays the primitives of \S\ref{sec:primitives} against running agents through their own order paths on paper-trading accounts, and reports each primitive's success per model and per safeguard configuration. Such a benchmark has to score the three properties of the final order defined in \S\ref{sec:threat} and the shift in net Greeks, because a decision-flip rate records no attack whenever the direction stays correct.
  \item \textbf{Provenance for derived statistics.} Just \nRtwoFull{} of the \nSchemes{} systems derive Greeks and IV from a pricing model or an authoritative chain with sanity checks (R2), and none cross-checks market data against an independent feed (C2). Although recomputing IV from the quoted bid and ask is cheap, text and tools also carry statistics such as IV rank and expected move that depend on history. An open question is whether such a number can carry a signed computation chain that an adaptive attacker cannot forge.
  \item \textbf{Leg-level atomicity.} A defined-risk order becomes naked when its short leg fills before its long leg, or when its long leg closes first. One purposive system already builds every spread as a single native multi-leg order (Appendix~\ref{app:schemes}). It remains open how an order-level verifier should bound the window between legs at brokers that route legs separately.
  \item \textbf{The price of unpredictability.} Randomized order timing and size, and refusing thin contracts, lower what a profile-and-harvest attacker (O-6) can collect. The size of that reduction and its execution cost to an honest agent are unmeasured, and together they decide whether these controls are worth deploying.
  \item \textbf{Adaptive attackers against platform bans.} A ban on uncovered writing leaves five of the six primitives reachable (\S\ref{sec:primitives}), and it bounds the Alpaca-centered population of \S\ref{sec:rq1}. An adaptive attacker~\cite{nasr2025attackermovessecond} who knows that a platform blocks O-1 may still reach a comparable loss through permitted structures, for example by stacking covered positions into concentrated assignment risk. It is also open whether defenses built for text-only equities agents survive such an attacker.
  \item \textbf{Keeping the measurement current.} \pctEligibleTwoWeek{} of the eligible repositories appeared within two weeks, and broker agent interfaces change their documentation monthly. Because a one-time snapshot ages quickly, the community needs a continuous measurement that reruns the screen and flags a broker that widens an agent's write scope, for which our screening script and scoring rules are a starting point (Appendix~\ref{app:search}).
\end{enumerate}

\section{Related Work}
\label{sec:related}

\begin{table}[t]
\centering
\footnotesize
\caption{This study compared with FARSIGHT~\cite{wang2026soktradingagentsmarket}, the equities systematization, dimension by dimension. The differences follow from the four extra decision variables of an options order and from measuring systems that run today.}
\label{tab:farsight}
\setlength{\tabcolsep}{3pt}
\renewcommand{\arraystretch}{1.1}
\begin{tabularx}{\columnwidth}{@{}>{\RaggedRight\arraybackslash}p{1.75cm} >{\RaggedRight\arraybackslash}X >{\RaggedRight\arraybackslash}X@{}}
\toprule
& \textbf{FARSIGHT (equities)} & \textbf{This study (options)} \\
\midrule
Attacker levers & 1: direction (plus size) & 5: direction, magnitude, time, volatility, structure (plus size) \\
Worst case & Wrong direction; unbounded through an explicit short & Unbounded loss from one dropped leg, direction unchanged; assignment into unplanned stock \\
Systems scored & 15 academic designs & \nSchemes{} systems (\nAcademic{} academic, \nBroker{} broker-native, \nOSS{} open-source), from a logged screen of \nUnion{} repositories \\
Deployment & Proposed in the literature & Live: agents place options orders at retail brokers \\
Codebook & Robustness (R1--R4, turbulence) and security (S1--S3) & Same two dimensions: 7 robustness and 11 security indicators, plus a profile and a computed impact \\
New dimensions & None & Write scope, Greeks and IV provenance, expiry and assignment, order-flow observability \\
Headline & 80\% fail a robustness metric; all 15 vulnerable & \nInputVerified{} of \nSchemes{} verify their inputs, \nWfour{} can reach an uncovered short, and \nHunbounded{} can reach unbounded loss \\
\bottomrule
\end{tabularx}
\end{table}

\rih{Security of LLM trading agents.}\label{sec:farsight} The closest work is FARSIGHT, the equities systematization of Wang and Saxena~\cite{wang2026soktradingagentsmarket}, which evaluated 15 academic equities designs and found that 80\% fail a core robustness metric and all 15 have security vulnerabilities. Its unit of harm is a flipped buy-or-sell decision on a linear position. TradeTrap perturbs single components of equities trading agents and reports extreme concentration and large drawdowns in closed-loop backtests~\cite{tradetrap2025}, and Market Signal Injection changes formatting, ordering, or commentary to shift LLM pricing-agent decisions in simulated oligopolies~\cite{marketsignalinjection2026}. OQL, the closest options system, translates natural-language strategy descriptions into a domain-specific language that an engine validates and instantiates deterministically, and it is evaluated for translation accuracy~\cite{oql2026}. Table~\ref{tab:farsight} compares our study with FARSIGHT dimension by dimension. In contrast, we measure the options agents that brokers and developers ship today and price each failure by the worst loss of the order it produces.

\rih{Agent-security benchmarks and systematizations.} AgentDojo~\cite{debenedetti2024agentdojo}, InjecAgent~\cite{zhan2024injecagent}, and the Agent Security Bench~\cite{zhang2025asb} measure whether injected content triggers unintended tool calls, in the indirect prompt-injection setting of Greshake et al.~\cite{greshake2023}, and MCPTox measures tool poisoning on real MCP servers~\cite{mcptox2026}. Kim et al.~\cite{kim2026sokagentic} systematize attacks and defenses across agentic AI systems, and a systematization of agentic commerce~\cite{agenticcommercesok2026} treats autonomous agents that transact on a user's behalf. Defenses such as CaMeL~\cite{camel2025}, spotlighting~\cite{spotlighting2024}, StruQ~\cite{struq2025}, and SecAlign~\cite{secalign2025} separate instructions from data. In contrast, we measure which protections deployed options agents enforce, and we score actions whose loss depends on the legs of an order and is counted in dollars at expiry.

\rih{Adversarial trading and order-flow leakage.} Trading against a victim whose orders leak is a known attack class. Flash Boys 2.0 documented front-running bots on decentralized exchanges~\cite{daian2020flashboys}, and a systematization of decentralized-finance attacks organized the class~\cite{zhou2023sokdefi}. In contrast, our counterparty primitive (O-6) targets the observable options orders of an LLM agent in thin listed contracts, where the attacker first induces the order that it then fills.

\section{Discussion}
\label{sec:discussion}

\rih{The broker as a point of control.} Because an agent's worst case is set by the broker's write policy and the account's options level (\S\ref{sec:rq1}, \S\ref{sec:rq2}), one broker decision protects every agent on the platform at once. Robinhood and Alpaca already exclude uncovered writing, and their agents stop at assignment whatever the agent code checks. The other brokers let an agent inherit the options level of the human who connected it. FINRA Rule 2360 approves that level by assessing the customer's experience and finances~\cite{finra2360}, which says nothing about the software that will place the orders. An agent-specific approval level that excludes uncovered writing would therefore remove the unbounded tail from \nBrokerUnbounded{} of the \nBroker{} broker interfaces we scored without changing a line of agent code.

\rih{Size limits and loss limits.} The pre-trade controls that market-access rules require are written as caps on size, notional value and credit~\cite{sec2010rule15c3_5}, and C8 is the safeguard we found most often. On options, however, a cap on contracts or notional leaves a naked short unbounded, and \nCeightFullUnbounded{} of the \nHunbounded{} systems that reach unbounded loss enforce such a cap. The control that bounds the induced order is a maximum loss computed from the legs before entry (R1), which \nRoneFullOrder{} of the \nOrderCapable{} order-capable systems enforce. Brokers that already compute margin per position hold the numbers this check needs.

\rih{Evaluating direction-preserving attacks.} Agent-security benchmarks count unintended tool calls~\cite{debenedetti2024agentdojo,zhan2024injecagent,zhang2025asb}, and trading-agent evaluations count flipped buy-or-sell decisions~\cite{wang2026soktradingagentsmarket}. An options attack that keeps the direction and drops a protective leg triggers neither, because the agent called the intended tool and chose the intended side. Evaluations of options agents should therefore score the risk class of the resulting order, the change in maximum loss between the honest and the induced order, rather than whether the decision changed, and no system we scored documents a full adversarial evaluation (C11).

\rih{A fast-turnover ecosystem.} Most of the open-source population appeared within two weeks around one event (\S\ref{sec:rq1}), and much of it is likely to stop being maintained once the event ends. Its designs nevertheless persist through forks and templates, and openalgo alone has 1{,}275 forks. Review effort is better spent on the shared layers that many agents reuse, namely broker SDKs, MCP servers and starter templates, than on individual repositories, because a missing coverage check in one shared order tool reaches every agent built on it.

\rih{Generality.} Index, futures and crypto options share all five decision variables, so the threat model and codebook apply to them unchanged, and our scored set already includes futures options (tastytrade) and crypto options (OKX, a Deribit agent). Although the enabling layer is deployed (\S\ref{sec:background}), we found no public report of a realized loss from an autonomous options agent, so the window for building controls into the order path is open now.

\rih{Limitations.} Three limitations bound how far the counts extend. First, we score documented designs at one snapshot, so the RQ2 and RQ3 counts describe what each paper, repository, or broker page commits to, and a running deployment may add or drop safeguards. Second, one coder and LLM-assisted passes produced the scores and we report no inter-coder agreement, so cells near the line between full and partial in Table~\ref{tab:schemes} carry one coder's judgment. Third, proprietary desks are absent, because our search finds public systems.

\section{Conclusion}
\label{sec:conclusion}

We measured the LLM options-trading agents that brokers and developers ship and scored each by the worst loss its order path can reach. Most of the open-source population appeared in 2026 around one hackathon. None of the \nSchemes{} systems verified their inputs and \nHunbounded{} could reach unbounded loss, so controls for options agents belong at the order boundary, outside the model.

\bibliographystyle{IEEEtran}
\bibliography{main}

\appendix

\section{Attack Classes That Reach the Primitives}
\label{app:migration}\label{sec:migration}

Table~\ref{tab:migration} in \S\ref{sec:primitives} lists nine publicly documented attack classes, grouped by the FARSIGHT attack taxonomy~\cite{wang2026soktradingagentsmarket}, that reach an options agent with the attacker's capabilities unchanged. A class enters the table if a peer-reviewed or public preprint exemplar demonstrates it on some LLM victim, if the attacker's action uses a surface of \S\ref{sec:threat}, and if its escalation follows from payoff mechanics~\cite{hull2021,blackscholes1973} and names at least one axis of \S\ref{sec:thesis}. The effect column reports what each exemplar measured on its own victims, none of which traded options, and the escalation column states what the same corruption costs on options. Each class feeds at least one primitive of Table~\ref{tab:primitives}, and the escalations fall into three patterns.

\rih{Direction-preserving loss.} On an equities agent, an information-source attack pays when it flips the buy-or-sell decision. Rows M1, M4, and M5 instead impose a loss while the agent picks the correct direction. An option priced on implied volatility and time (A4) loses value when a move arrives late or smaller than the premium implies, so inflating the expected size of a move sells the agent an overpriced bet in the right direction, and a wide spread (A3) makes a manufactured flow narrative cost the spread on entry and exit.

\rih{New order fields.} Rows M7 to M9 act on the order itself. An equity order has a side, a size, and a price, while an options order adds legs, ratios, strikes, expiries, an opening or closing flag, and exercise decisions (A5). An injected instruction or a poisoned tool response therefore has more fields to hide in, and a single altered leg changes the risk class of the whole position (A1) while the order still looks routine. On an options account, the falsified state field of TradeTrap's M8 result~\cite{tradetrap2025} would be a margin requirement or a maximum-loss figure.

\rih{Framing without falsehood.} Row M3 matters most for O-1, because writing an option collects a premium kept in most outcomes and gives back a large multiple of it in the rare adverse one~\cite{hull2021}. Instruction-tuned models exhibit the certainty effect~\cite{itzhak2024instructed}, which overweights outcomes presented as near-certain~\cite{tversky1981framing}, and ``this put expires worthless in most months'' can be literally true while pointing the agent toward the payoff that hides its tail. A fact-check finds nothing to reject, because each stated fact checks out.

\section{Safeguard Codebook}
\label{app:codebook}

Table~\ref{tab:codebook-full} defines every indicator, the rule that separates a full score from a partial one, and the public framework or axis the indicator operationalizes. For every robustness and security indicator, \halfc{} applies when the safeguard is documented but depends on the model following an instruction, is optional or off by default, or can be cleared by the agent itself, and \emptyc{} applies when the public artifacts document nothing. Scores are design-level, from public artifacts at the snapshot, with cell-level evidence in Appendix~\ref{app:schemes}. Five rules make the profile and impact cells mechanical, and a consistency check rejects any cell that violates them. (1) X2 and X3 describe the most permissive configuration that the documentation or code lets an operator enable, and X3 counts the broker's own restrictions and the account's highest options level. (2) A gate yields A2 only if it applies to every order in every documented configuration. A gate that the agent can clear, that is off by default, or that a documented setting removes yields A3. (3) C8 is full when at least one control outside the model that the agent cannot clear applies to every order, partial when such controls are optional or agent-clearable or when only broker eligibility rules apply, and none otherwise. C8 and C10 do not apply to a system without an order path (A0 or W0). (4) H1 is none for A0, A1, or W0 and otherwise premium (W1), assignment (W2, W3), or unbounded (W4), and H2 is platform for broker-native and adopted open-source systems. (5) H3 is interceptable for A0 to A2, partial for A3 or when C8 is full or partial, and not interceptable otherwise. One coder scored the systems of the initial snapshot, and the \nAddedSchemes{} systems added at this snapshot were scored with LLM assistance and re-checked against their sources in a second LLM-assisted pass. Every cell changed by these rules keeps its previous value in the evidence file. We have not run a second-coder pass and report no inter-coder agreement.

\begin{table*}[t]
\centering
\footnotesize
\caption{The safeguard codebook in full, in the order of Figure~\ref{fig:codebook}: robustness (R), security (C), profile (X), and computed impact (H). The Grounding column names the amplification axis (\S\ref{sec:thesis}) or public framework the indicator operationalizes: the OWASP Top 10 for LLM Applications~\cite{owasp2025llm}, SEC Rule 15c3-5 on market-access risk controls~\cite{sec2010rule15c3_5}, and FINRA Rule 2360 options approval levels~\cite{finra2360}.}
\label{tab:codebook-full}
\setlength{\tabcolsep}{3pt}
\renewcommand{\arraystretch}{1.04}
\begin{tabularx}{\textwidth}{@{}l>{\RaggedRight\arraybackslash}p{3.1cm}>{\RaggedRight\arraybackslash}X>{\RaggedRight\arraybackslash}p{3.0cm}@{}}
\toprule
\textbf{ID} & \textbf{Indicator} & \textbf{Levels or rule for \fullc{}} & \textbf{Grounding} \\
\midrule
R1 & Maximum loss & Maximum loss is computed and enforced before entry, distinguishing defined from undefined risk. & A1 \\
R2 & Greeks and IV provenance & Greeks and IV are computed from a pricing model or an authoritative chain, with sanity checks, not read from text. & A2 \\
R3 & Spread and liquidity & Execution enforces limit orders, open-interest or volume floors, or maximum spreads. & A3, A6 \\
R4 & Decay and expiry & Time decay, event volatility, and days to expiry gate entries and exits. & A4 \\
R5 & Exercise and assignment & Early assignment, exercise, and pin risk are handled by rule. & A5 \\
R6 & Stress & Volatility spikes, gaps, halts, and stale data trigger enforced behavior. & Turbulence dimension of~\cite{wang2026soktradingagentsmarket} \\
R7 & Portfolio limits & Position size, net Greeks, concentration, or loss limits are enforced. & \cite{wang2026soktradingagentsmarket} \\
\midrule
C1 & Source authentication & Text inputs are authenticated or provenance-checked before use. & OWASP LLM04 data poisoning \\
C2 & Market-data integrity & Quotes, chains, or Greeks are cross-validated against an independent authoritative feed. & OWASP LLM04 \\
C3 & Tool and MCP trust & Servers are allowlisted or pinned and tool outputs are validated. & OWASP LLM03 supply chain; \cite{mcptox2026} \\
C4 & Instruction--data separation & Untrusted content is structurally separated from instructions, or injection is mitigated in code. & OWASP LLM01 prompt injection; \cite{greshake2023} \\
C5 & Memory and retrieval & Stored memory and retrieved documents are validated before reuse. & OWASP LLM08 vector and embedding weaknesses \\
C6 & Multi-agent trust & Messages between agents are authenticated or constrained; single-agent systems score \emptyc{} with a note. & Attack taxonomy of~\cite{wang2026soktradingagentsmarket} \\
C7 & Supply chain & Configuration, prompts, models, or dependencies are pinned or provenance-checked. & OWASP LLM03 \\
C8 & Out-of-model pre-trade limits & A control outside the LLM that the agent cannot clear applies to every order: a size, notional, or loss cap, or a mandatory human confirmation. Not applicable without an order path. & SEC Rule 15c3-5 \\
C9 & Least privilege & Dedicated agent account, scoped options level, capped buying power, or scoped keys. & OWASP LLM06 \\
C10 & Monitoring and kill switch & Logging, alerting, and a kill switch are enforced. Not applicable without an order path. & SEC Rule 15c3-5 \\
C11 & Adversarial evaluation & Red-teaming, injection tests, or adversarial backtests are documented. & This work (\S\ref{sec:recommend}) \\
\midrule
X1 & System type & Academic system, open-source project, or broker-native agent interface. & Descriptive \\
X2 & Autonomy & A0 advisory, no orders; A1 the LLM can only veto; A2 a human approves every order in every documented configuration; A3 automatic behind a gate the agent can clear, that is off by default, or that a documented setting removes; A4 no gate. & OWASP LLM06 excessive agency \\
X3 & Write scope & Maximum reachable through the documented order path, counting broker restrictions and the account's highest options level: W0 read-only; W1 long premium; W2 defined-risk multi-leg; W3 covered or cash-secured shorts; W4 uncovered shorts. & FINRA Rule 2360 approval levels \\
X4 & Instruments & Equity, index, 0DTE, crypto, or futures options; multi-leg orders. & Descriptive \\
X5 & Input channels & I1 public text; I2 market data, chains, Greeks; I3 tool or MCP output; I4 memory or retrieval; I5 peer agents; I6 operator configuration, prompts, model, dependencies. & Threat model (\S\ref{sec:threat}) \\
\midrule
H1 & Worst reachable loss & None for A0, A1, or W0; otherwise premium (W1), assignment (W2, W3), or unbounded (W4). & O-1 (\S\ref{sec:primitives}) \\
H2 & Blast radius & One user, one design deployed to many users (platform), or the market. Platform for broker-native and adopted open-source systems. & Descriptive \\
H3 & Interceptability & Interceptable for A0 to A2; partial for A3 or when C8 is full or partial; not interceptable otherwise. & X2, C8 \\
\bottomrule
\end{tabularx}
\end{table*}

\section{Per-System Evidence}
\label{app:schemes}

Each block below covers one scored system: the public artifact behind it, the evidence for its autonomy (X2) and write-scope (X3) levels, the derivation of its worst reachable loss (H1), and which codebook indicators (Appendix~\ref{app:codebook}) it scores full, partial, or none. Quotations are shortened to the evidence needed to justify a score. Of the \nExamined{} systems examined in depth, \nExcludedReview{} were excluded after review. That list, with its reasons, closes the appendix.

\rih{1. Kuang and Lin~\cite{kuanglin2025wheel}.} arXiv preprint v1 only (2025, q-fin.CP). No later version, code repository, or data-availability statement found. Autonomy A0: The paper contains no broker API, order-execution routine, or live or paper-trading infrastructure. It is described as purely a backtest and simulation study. Write scope W0: No order-writing capability of any kind is described anywhere in the paper. The system outputs a recommendation a human would execute manually. Worst reachable loss (none): X2 is A0 and X3 is W0, so the system describes no order-writing capability. A human who followed its output manually could still lose money. \emph{Full:} none. \emph{Partial:} R3, R4, R5, R6, R7, C5, C6, C7. \emph{None:} R1, R2, C1, C2, C3, C4, C9, C11.

\rih{2. OQL~\cite{oql2026}.} arXiv preprint v1 only (2026, cs.AI). No code, data, or artifact link on the abstract or HTML page. Autonomy A0: The output is a ranked set of backtested candidate strategies. No order placement, broker, paper-trading, or live-execution component is described anywhere in the paper. Write scope W0: No broker API or order-writing capability is described. The engine only reads option-chain data and returns contract combinations for a human to consider. Worst reachable loss (none): X2 is A0 and X3 is W0, so the system itself can cause no automated loss. A human who manually executed its output could still lose money. \emph{Full:} none. \emph{Partial:} R1, R2, R4. \emph{None:} R3, R5, R6, R7, C1, C2, C3, C4, C5, C6, C7, C9, C11.

\rih{3. OKX Agent Trade Kit~\cite{repo_okxagenttradekit}.} Official server, 437 stars, 68 forks, MIT, active. Repo created in 2026. Autonomy A4: option\_place\_order posts directly to the exchange's order endpoint with no confirmation, dry-run, or preview parameter anywhere in the handler. Write scope W4: Writing an option is a margined position by product design. There is no covered or cash-secured order type distinct from an ordinary margined sell. Worst reachable loss (unbounded): No code or documentation computes or enforces a maximum-loss figure, and nothing outside the model can stop a manipulated option order before it executes. \emph{Full:} none. \emph{Partial:} R2, C3, C7, C9, C10. \emph{None:} R1, R3, R4, R5, R6, R7, C1, C2, C4, C5, C6, C8, C11.

\rih{4. Public MCP~\cite{repo_publicmcp}.} Official server, 66 stars, 7 forks, Apache-2.0, active. Repo created in 2026. Autonomy A4: Each order tool builds a request from raw arguments and immediately calls the SDK's place method. No confirmation token or dry-run step links a preview call to it. Write scope W4: place\_order and place\_multileg\_order accept sell-to-open legs with only syntactic validation. No coverage, margin, or naked-option check exists anywhere in the file. Worst reachable loss (unbounded): No gate exists before execution, uncovered short options are reachable with no coverage check, and no cap of any kind exists in code. \emph{Full:} none. \emph{Partial:} R2, R3, C3, C7, C9, C10. \emph{None:} R1, R4, R5, R6, R7, C1, C2, C4, C5, C6, C8, C11.

\rih{5. Webull MCP~\cite{repo_webullmcp}.} Official server, 25 stars, 4 forks, Apache-2.0, active. Repo created in 2026. Autonomy A4: Order tools call the syntactic validator, then the SDK's place\_order directly. No confirmation token or required preview links to the place tools. Write scope W4: Strategy validators permit straddles and strangles with no side or coverage constraint, so a nominal strategy order can be an uncovered short position. Worst reachable loss (unbounded): The advertised notional and quantity risk controls exist in code but are never called from either option-order validator, so nothing caps an option order. \emph{Full:} none. \emph{Partial:} R2, C3, C7, C9, C10. \emph{None:} R1, R3, R4, R5, R6, R7, C1, C2, C4, C5, C6, C8, C11.

\rih{6. tastytrade MCP~\cite{tastytrademcp2026}.} Active, official server, 20 stars, 7 forks. Repo created in 2026. Autonomy A3: The only built-in order gate is a dry-run confirmation token that the same agent both mints and redeems, so the agent can clear it itself. Write scope W4: Order legs accept Sell to Open for equity and future options with no local covered-or-uncovered check, so maximum authority equals the account's options level. Worst reachable loss (unbounded): Uncovered Sell to Open is reachable whenever the account's options level allows it, and the agent can clear the only order-level gate itself. \emph{Full:} C3, C7, C8. \emph{Partial:} R1, R2, R3, R4, R5, R6, R7, C1, C4, C5, C9, C10, C11. \emph{None:} C2, C6.

\rih{7. IBKR AI integration~\cite{ibkraiintegration2026}.} Hosted broker product with no source code or repository. Scored from the official product page and FAQ (direct fetch blocked, browser used). Autonomy A2: The AI drafts order instructions, including options strategies, and a human reviews and submits each one in the AI Instructions tab. Write scope W4: Instructions can express any options order the account is approved for, and IBKR Level 4 permits naked calls, straddles, and strangles. Worst reachable loss (unbounded): A drafted uncovered short that a human approves reaches unbounded loss. \emph{Full:} C8. \emph{Partial:} C9. \emph{None:} R1, R2, R3, R4, R5, R6, R7, C1, C2, C3, C4, C5, C6, C7, C10, C11.

\rih{8. TradeStation MCP~\cite{tradestationmcp2026}.} Hosted connector with no public repository or tool schema. Scored from two official TradeStation pages at the snapshot. Autonomy A2: Both official pages state confirmation is mandatory and unconditional: no trade executes without the user's explicit approval before it is sent. Write scope W4: The connector claims it never narrows the account's own trading permissions, and TradeStation's general options-approval scale documents a tier that permits uncovered writing. Worst reachable loss (unbounded): The connector inherits the account's full approval, which reaches uncovered writing, and confirmation is a generic per-order approval rather than a coverage check. \emph{Full:} C8. \emph{Partial:} C9, C10. \emph{None:} R1, R2, R3, R4, R5, R6, R7, C1, C2, C3, C4, C5, C6, C7, C11.

\rih{9. Alpaca MCP~\cite{repo_alpacamcp}.} Official server, 991 stars, 317 forks, MIT, active. Repo created in 2025. Autonomy A4: place\_option\_order posts directly to the order endpoint with no confirmation token, preview tool, or dry-run step anywhere in the code. Write scope W3: The server code has no local coverage check, but Alpaca's brokerage platform independently rejects uncovered options at every options level, regardless of tier. Worst reachable loss (assignment): Covered calls, cash-secured puts, and defined-risk spreads are reachable under full automation, but the platform rejects uncovered options outright at any level. \emph{Full:} R1. \emph{Partial:} R2, R3, R5, C1, C3, C4, C7, C8, C9, C11. \emph{None:} R4, R6, R7, C2, C5, C6, C10.

\rih{10. Robinhood agentic options~\cite{robinhoodoptions2026,robinhoodmcp2026}.} Production broker feature, labeled beta. Agentic Trading launched in May 2026 for equities, with options added in a later rollout. Autonomy A3: The only gate before execution is a review step the agent or user can skip. The support article confirms trades can execute without per-order confirmation. Write scope W3: The order tool documents Level 2 strategies (covered calls, cash-secured puts) and Level 3 multi-leg spreads. Naked options and margin borrowing are not offered. Worst reachable loss (assignment): Covered calls and cash-secured puts carry assignment risk. Naked options are banned platform-wide, so the documented order path cannot reach unbounded loss. \emph{Full:} R1, C9, C10. \emph{Partial:} R2, R3, R4, R5, R6, R7, C8. \emph{None:} C1, C2, C3, C4, C5, C6, C7, C11.

\rih{11. lumibot~\cite{repo_lumibot}.} Active open-source project (GPL-3.0), \starsLumibot{} GitHub stars and 404 forks at the snapshot. Autonomy A4: allow\_trading defaults to True, and the agent submits orders directly to the broker through submit\_order with no per-order human approval. Write scope W4: Option orders accept side sell\_to\_open with no coverage check in code. Only chain-read readiness and closing-side consistency are verified. Worst reachable loss (unbounded): Sell-to-open single-leg and arbitrary multi-leg option orders are accepted with no coverage or margin check anywhere in the codebase. \emph{Full:} none. \emph{Partial:} R1, R2, R3, R4, R5, R6, R7, C1, C3, C4, C5, C6, C7, C8, C9, C10, C11. \emph{None:} C2.

\rih{12. tasty-agent~\cite{repo_tastyagent}.} Active, community/unofficial MCP server, 86 stars, 26 forks, MIT. Created in 2024. No test-gated CI. Autonomy A4: place\_order executes immediately on a single call with dry\_run defaulting to False. The only 'require confirmation' rule lives in a skill prompt, not in code. Write scope W4: Sell-to-open option legs are accepted with only an instrument-category check, no coverage check anywhere in the order path. Worst reachable loss (unbounded): A single tool call can submit an uncovered short option order whenever the account's options level allows it, with no cap or confirmation the agent cannot clear. \emph{Full:} none. \emph{Partial:} R1, R2, R3, R4, R6, C3, C7, C10. \emph{None:} R5, R7, C1, C2, C4, C5, C6, C8, C9, C11.

\rih{13. openalgo~\cite{repo_openalgo}.} \starsOpenalgo{} GitHub stars, 1,275 forks, AGPL-3.0, active. Autonomy A3: The MCP path auto-executes: every MCP order tool calls the broker API directly, with no approval step placed in between. Write scope W4: place\_options\_order runs no cover or margin check before dispatch, so uncovered short calls and puts on multiple exchanges can be reached. Worst reachable loss (unbounded): The MCP path auto-executes and naked short options are reachable. The only pre-trade limit found is partial, not a hard cap. \emph{Full:} R2, C10. \emph{Partial:} R1, R3, R4, R6, R7, C1, C3, C4, C5, C7, C8, C9. \emph{None:} R5, C2, C6, C11.

\rih{14. schwab-mcp~\cite{repo_schwabmcp}.} Community server, 61 stars, 19 forks, MIT, active. Repo created in 2025. Autonomy A3: Discord approval gates each order, but the documented --jesus-take-the-wheel flag removes the gate. Write scope W4: Sell-to-open legs are accepted with no coverage check anywhere in the code, so authority equals the account's real options approval level. Worst reachable loss (unbounded): Uncovered orders are reachable, and the only non-model control, Discord approval, is fully bypassable through a documented flag with no max-loss check. \emph{Full:} none. \emph{Partial:} R2, C7, C8, C9, C10. \emph{None:} R1, R3, R4, R5, R6, R7, C1, C2, C3, C4, C5, C6, C11.

\rih{15. trade-it-mcp~\cite{repo_tradeitmcp}.} Hosted, closed-source multi-broker service. GitHub repo is documentation/registry-metadata only (60 stars, 15 forks, CC0-1.0). Scored from this repo plus docs.tradeit.app at the snapshot. Autonomy A3: A draft-then-execute confirmation exists, but the agent itself decides when to call execute\_trade. Write scope W4: The leg schema allows a single sell-to-open leg with no documented pairing or coverage requirement, so authority equals the linked brokerage's own options-approval tier. Worst reachable loss (unbounded): A single sell-to-open leg is reachable with no documented coverage requirement, and nothing documented caps the loss or blocks execution outside the model. \emph{Full:} none. \emph{Partial:} C8, C9, C10. \emph{None:} R1, R2, R3, R4, R5, R6, R7, C1, C2, C3, C4, C5, C6, C7, C11.

\rih{16. OpenProphet~\cite{repo_openprophet}.} 112 stars, 39 forks, CC BY-NC 4.0. Created in 2026. Featured in Bloomberg. Autonomy A4: The heartbeat loop executes trades without human intervention, and every guardrail lives in one object the agent can overwrite through a single call. Write scope W3: Sell-to-open is accepted in code, but Alpaca, the only broker, rejects uncovered options at every level. Worst reachable loss (assignment): Covered and cash-secured shorts on Alpaca bound the worst case at assignment. \emph{Full:} none. \emph{Partial:} R2, R3, R4, R6, R7, C3, C8, C10. \emph{None:} R1, R5, C1, C2, C4, C5, C6, C7, C9, C11.

\rih{17. trading\_skills~\cite{repo_tradingskills}.} 368 stars, 85 forks, MIT, active. Created in 2026. Positioned for Claude Desktop. Autonomy A3: The live-order tool defaults to a dry run, but execute is a plain boolean the calling model sets itself, with no separate human-confirmation step. Write scope W2: Executing places the chosen spread as a single native combo order margined as one defined-risk position. Only verticals and iron condors are ever opened. Worst reachable loss (assignment): Short legs of the vertical and iron-condor spreads carry assignment exposure even though the combo order's structure keeps the position's maximum loss capped. \emph{Full:} R3, C8. \emph{Partial:} R1, R2, R4, R5, R6, R7, C7, C9, C10. \emph{None:} C1, C2, C3, C4, C5, C6, C11.

\rih{18. TastyScanner MCP~\cite{repo_tastyscanner}.} 3 stars, 1 fork, MIT, active. Repo created in 2026. Autonomy A4: Once ENABLE\_LIVE\_TRADING is set, execute\_trade passes the agent's legs straight to the broker's order service with no per-order confirmation or preview step. Write scope W4: The execute\_trade leg schema accepts any Sell to Open leg with an unbounded quantity and no coverage or margin check anywhere in the server. Worst reachable loss (unbounded): The strategy builder proposes naked short structures, and execute\_trade accepts arbitrary uncovered Sell to Open legs with no quantity cap. \emph{Full:} none. \emph{Partial:} R1, R2, R3, R4, R6, C3, C7, C9, C10. \emph{None:} R5, R7, C1, C2, C4, C5, C6, C8, C11.

\rih{19. deribit agent (Vida)~\cite{repo_deribitagent}.} 4 stars, MIT, not archived. A personal, self-hosted terminal. Autonomy A2: Every gated write tool registers a confirmation request, and the loop waits up to 30 seconds for a human response before proceeding. Write scope W4: place\_order accepts any instrument, including options, with direction sell or buy and no check anywhere for covered or defined risk. Worst reachable loss (unbounded): Naked option shorts and leveraged perpetuals are reachable through place\_order. Human confirmation lowers but does not eliminate the risk of a bad order. \emph{Full:} C8. \emph{Partial:} R1, R2, R3, R4, R6, R7, C3, C7, C9, C10. \emph{None:} R5, C1, C2, C4, C5, C6, C11.

\rih{20. options-trading-bot~\cite{repo_rohan5commit}.} 0 stars, 44 commits, not archived. Created in 2026. Autonomy A4: A scheduled workflow runs the full pipeline from LLM decision to order submission with no human approval step anywhere. Write scope W3: The validator accepts any leg with side buy or sell and positive quantity, without checking that leg sides match the named strategy. Worst reachable loss (assignment): Single-leg sell puts are reachable as cash-secured shorts, and spread short legs carry assignment risk under full automation with no max-loss check. \emph{Full:} C8, C9. \emph{Partial:} R1, R2, R3, R4, R5, R6, R7, C5, C10. \emph{None:} C1, C2, C3, C4, C6, C7, C11.

\rih{21. Agentic-Trader~\cite{repo_agentictrader}.} 3 stars, no license, not archived. Described as a sanitized public fork. Autonomy A3: Auto-approval within gates is enabled by default in configuration, so gate-clean proposals are submitted automatically and only posted to Slack afterward. Write scope W3: Allowed structures are long calls and puts, debit spreads, and cash-secured puts. A cash-secured put requires several settings to all be enabled. Worst reachable loss (assignment): A cash-secured put is reachable when credit entries are enabled. Its loss is bounded by the strike times 100 minus the premium received. \emph{Full:} R1, R2, R3, R7, C3, C6, C8, C10. \emph{Partial:} R4, R5, R6, C1, C2, C4, C5, C7, C9, C11. \emph{None:} none.

\rih{22. agentic-trading-research~\cite{repo_agentictradingresearch}.} 11 stars, GPL-3.0, not archived. Described as an LLM-based options trader. Autonomy A4: The scheduled trading cycle runs automatically at a fixed interval while the market is open, with no human approval step. Write scope W1: The order gateway hard-codes buy-side market orders for opening. The sell path only closes existing long positions and rejects oversized sell quantities. Worst reachable loss (premium): Only long calls and puts can be opened, with code-enforced buying-power caps, so the worst-case loss is bounded by the premium paid. \emph{Full:} R4, C8. \emph{Partial:} R1, R2, R3, R5, R6, R7, C1, C6, C9, C10. \emph{None:} C2, C3, C4, C5, C7, C11.

\rih{23. alpaca-options-agent~\cite{repo_alpacaoptionsagent}.} Hackathon submission, 0 stars, 0 forks, MIT, active. Created in 2026. A judged paper account only. Autonomy A1: A deterministic risk gate runs before the LLM and rejects any failing candidate. The LLM can only decline a gate-approved candidate, never open a rejected one. Write scope W2: Every open or close call builds a single native multi-leg order, never two independent legs, so the spread structure cannot leave an uncovered position. Worst reachable loss (none): X2 is A1, so per the codebook's rule the LLM cannot itself cause an order. A deterministic gate, not the LLM, fully specifies and submits the trade. \emph{Full:} R1, R2, R3, R4, R7, C8, C9, C10. \emph{Partial:} R5, R6, C3, C7. \emph{None:} C1, C2, C4, C5, C6, C11.

\rih{24. Theta Desk~\cite{repo_thetadesk}.} 1 star, 0 forks, no license. Created in 2026. Active. Autonomy A1: A ten-check code gate must pass before the LLM is ever called, and the LLM is then used only as a one-time veto. Write scope W4: The strategy registry includes naked short straddles and strangles at the money, which is the code's default, alongside defined-risk spreads. Worst reachable loss (none): The bot places no real orders. All trading runs against virtual paper capital, so no automated loss can reach a real account. \emph{Full:} R4, R7, C8, C10. \emph{Partial:} R1, R2, R3, R5, R6, C5, C7, C9. \emph{None:} C1, C2, C3, C4, C6, C11.

\rih{25. AI-trading-Agent~\cite{repo_liyang}.} 1 star, 0 forks, no license, not archived. Created in 2025. Autonomy A0: The only output is a recommendations file and a summary. No broker or order-placement code exists anywhere in the repository. Write scope W0: There is no write path to any broker. The compliance documentation states trading-account credentials are never collected, stored, or transmitted. Worst reachable loss (none): With no broker integration, credentials, or order code, a manipulated recommendation cannot lose money without a human acting on it first. \emph{Full:} none. \emph{Partial:} R4, R6, R7, C1, C5, C6, C9. \emph{None:} R1, R2, R3, R5, C2, C3, C4, C7, C11.

\rih{26. news-agent~\cite{repo_newsagent}.} 1 star, 0 forks, MIT. Created in 2026. Public as of the snapshot. Autonomy A0: The system only writes briefing files and sends an email summary. No order-placement code path exists anywhere in the repository. Write scope W0: Every write action is a file or an email. A repository-wide search for broker, order, or execute code finds nothing relevant. Worst reachable loss (none): The system has no order authority. Any loss depends entirely on a human choosing to act on a manipulated briefing. \emph{Full:} none. \emph{Partial:} R1, R2, R4, R6, C1, C7, C9. \emph{None:} R3, R5, R7, C2, C3, C4, C5, C6, C11.

\rih{Excluded after review.} \texttt{alpha-agent}: EXCLUDE. No LLM makes, recommends, or gates an options decision: a sentiment LLM feeds only ETF buy-and-hold scoring, a separate causal LM predicts stock prices, and the options-strategy code is deterministic, disconnected, and never called. \texttt{schwagent}: EXCLUDE. Options trading is fully automated, but every options decision is deterministic Python (theta wheel, covered-call screener, gamma scanner). The optional LLM only scores ETF sizing, writes filing summaries, or critiques backtests, never options.

\section{Search Protocol}
\label{app:search}

\rih{Snapshot.}
Searches, star counts, READMEs, and documentation pages reflect the snapshot. Repositories and broker documentation change monthly, so the scores describe that snapshot.

\rih{Sources and query strings.}
(i) \emph{Academic}: arXiv and security, machine-learning, and finance proceedings, with the terms ``options trading'', ``option strategy'', ``implied volatility'', ``Greeks'', and ``derivatives'' each combined with ``large language model'', ``LLM'', or ``agent'', plus backward and forward references from the equities SoK~\cite{wang2026soktradingagentsmarket} and recent trading-agent work~\cite{agentictrading2026,canllmstrade2025,tradetrap2025,marketsignalinjection2026}. (ii) \emph{GitHub}: the repository search API with six strings, \texttt{options trading LLM}, \texttt{option greeks llm}, \texttt{llm options trader}, \texttt{options trading agent}, \texttt{options trading ai agent}, and \texttt{options mcp server}, which returned \nQueryRaw{} hits in total. Every query stayed under the API's 1,000-result cap, and the union of all six holds \nUnion{} unique repositories. (iii) \emph{Broker documentation}: first-party agent interfaces, seeded by an industry test that found first-party agent MCP servers at 7 of 14 brokers, all launched in 2026~\cite{stockbrokers2026}.

\rih{Screening.} Figure~\ref{fig:funnel} in \S\ref{sec:method} summarizes the GitHub screen, and this paragraph gives its details. An automated keyword filter runs first: a repository survives if its name, description, topics, and README (first 30\,kB) match at least one of 21 LLM- or agent-related terms (e.g., ``llm'', ``mcp'', ``agentic'', or a named model family such as ``claude'' or ``gemini'') and at least one of 29 options-related terms (e.g., ``greeks'', ``iron condor'', ``0dte'', ``covered call''), case-insensitively. Both lists are recall-oriented by design: a bare ``expiry'' or ``mcp'' can fire on unrelated software, so this stage over-includes on purpose and the reading pass below removes false positives. Of the \nUnion{} candidates, \nStageOneKept{} survive for reading against four inclusion criteria: a candidate is eligible if (I1) an LLM makes, proposes, or gates an options decision (strategy, strike, expiry, structure, entry, or exit), or an interface is published for an LLM agent to do so, (I2) the system reads option-chain or option-position data, live or historical, (I3) a public paper, README, or documentation page supports scoring the candidate against the codebook (Appendix~\ref{app:codebook}), and (I4) the artifact is available on the snapshot date.

\rih{Exclusion criteria.}
Every other Stage 1 survivor is coded with one reason. (E1) LLM trading framework without options support (\nExcludeEone{}): the most-starred general frameworks we checked, \texttt{TradingAgents} (108,459 stars), \texttt{ai-hedge-fund} (63,736), \texttt{FinGPT} (21,288), and \texttt{FinRobot} (8,080), mention no option chains, Greeks, IV, strikes, or expiries in their READMEs. (E2) Options bot with no LLM in the decision path (\nExcludeEtwo{}). (E3) Public documentation insufficient to score (\nExcludeEthree{}). (E4) Attack, simulation, or taxonomy work rather than an options agent: no GitHub repository was excluded for this reason (\nExcludeEfour{}), but the same code separately removes TradeTrap~\cite{tradetrap2025} and Market Signal Injection~\cite{marketsignalinjection2026}, which study equities trading and pricing agents, the agentic-commerce SoK~\cite{agenticcommercesok2026}, which has no options-specific treatment, and a market-simulation study of LLM traders~\cite{canllmstrade2025}, all found through the academic search rather than GitHub. (E5) Not a trading system: education, data tools, or libraries without an LLM decision path (\nExcludeEfive{}). (E6) Duplicate, fork, or mirror of an included project (\nExcludeEsix{}). (E7) Empty or placeholder repository (\nExcludeEseven{}).

\rih{Outcome.}
Academic search surfaced two systems meeting I1 to I4, OQL~\cite{oql2026} and Kuang and Lin~\cite{kuanglin2025wheel}, and we scored both. Broker documentation and the industry test yielded \nBroker{} official interfaces: Robinhood~\cite{robinhoodoptions2026,robinhoodmcp2026} and tastytrade~\cite{tastytrademcp2026}, plus six more: OKX Agent Trade Kit~\cite{repo_okxagenttradekit}, Public MCP~\cite{repo_publicmcp}, Webull MCP~\cite{repo_webullmcp}, TradeStation MCP~\cite{tradestationmcp2026}, Alpaca MCP~\cite{repo_alpacamcp}, and IBKR AI integration~\cite{ibkraiintegration2026}. Seven scored repositories entered from outside the six-query union, through broker documentation or through searches outside the six strings: openalgo~\cite{repo_openalgo}, Theta Desk~\cite{repo_thetadesk}, tastytrade MCP, schwab-mcp~\cite{repo_schwabmcp}, trading\_skills~\cite{repo_tradingskills}, Public MCP, and Webull MCP. Beyond the scored set, \nEligibleUnscored{} eligible repositories remain unscored. Of the \nEligible{} eligible repositories, \nEligibleZeroStar{} have no stars, and \nEligibleTwoWeek{} were created in the two weeks around the hackathon. Appendix~\ref{app:search} lists the query strings.

\end{document}